\documentstyle[multicol,aps,prb,tabularx,epsf]{revtex}

\begin{document}
\title{Modeling of Tunneling Spectroscopy  in HTSC}
\author{Yu. M. Shukrinov~$^{a,b}$, A. Namiranian~$^{a}$,A. Najafi~$^{a}$}
\address{$^{a}$ Institute for Advanced Studies in Basic Sciences, Gava Zang,
Zanjan   45195-159, Iran\\
$^{b}$Physical Technical Institute of Tajik Academy of
Sciences\\
 299/1 Aini Str., Dushanbe, 734063, Tajikistan, C.I.S.}
\date{\today}
\maketitle
\begin{abstract}
The tunneling density of states of HTSC is calculated taking
into account tight binding band structure, group
velocity, and tunneling directionality for s-wave and d-wave gap symmetry.
The characteristic density of states has quasiparticle peaks' asymmetry, flat
s-wave and cusplike d-wave case subgap behavior, and asymmetric background.
We consider that the underlying asymmetry of the conductance peaks
is primarily due to the features of quasiparticle energy spectrum, and the
d-wave symmetry enhances the degree of the peaks' asymmetry. Increasing of the
lifetime broadening factor changes the degree of tunneling conductance peaks'
asymmetry, and leads to the confluence of the quasiparticle and van Hove singularity
peaks.

\end{abstract}
\begin{multicols}{2}

\section{ INTRODUCTION}
Tunneling measurements on HTSC have revealed a rich variety of properties
and characteristics [1-4]. They may be classified according to their low and high energy
features. With the low energy features we may attribute: (i) variable subgap
shape of conductance, ranging from sharp, cusplike, to a flat, BCS-like
feature [1]; (ii) voltage and temperature dependence of quasiparticle
conductivity [5,6]; (iii) subgap structure [2]; (iv) zero bias conductance
peak (ZBCP) [7]. The high energy features include: (i) asymmetry of conductance
peaks [1],(ii) van Hove singularity (VHS), (iii) conductance shape outside of the gap region (background (BG)) and
its asymmetry [1]; (iv) dip feature [8]; (v) hump feature [8].These features are collected in schematic Fig 1 .

\begin{figure}
\epsfysize=2.0in
\centerline{\epsffile{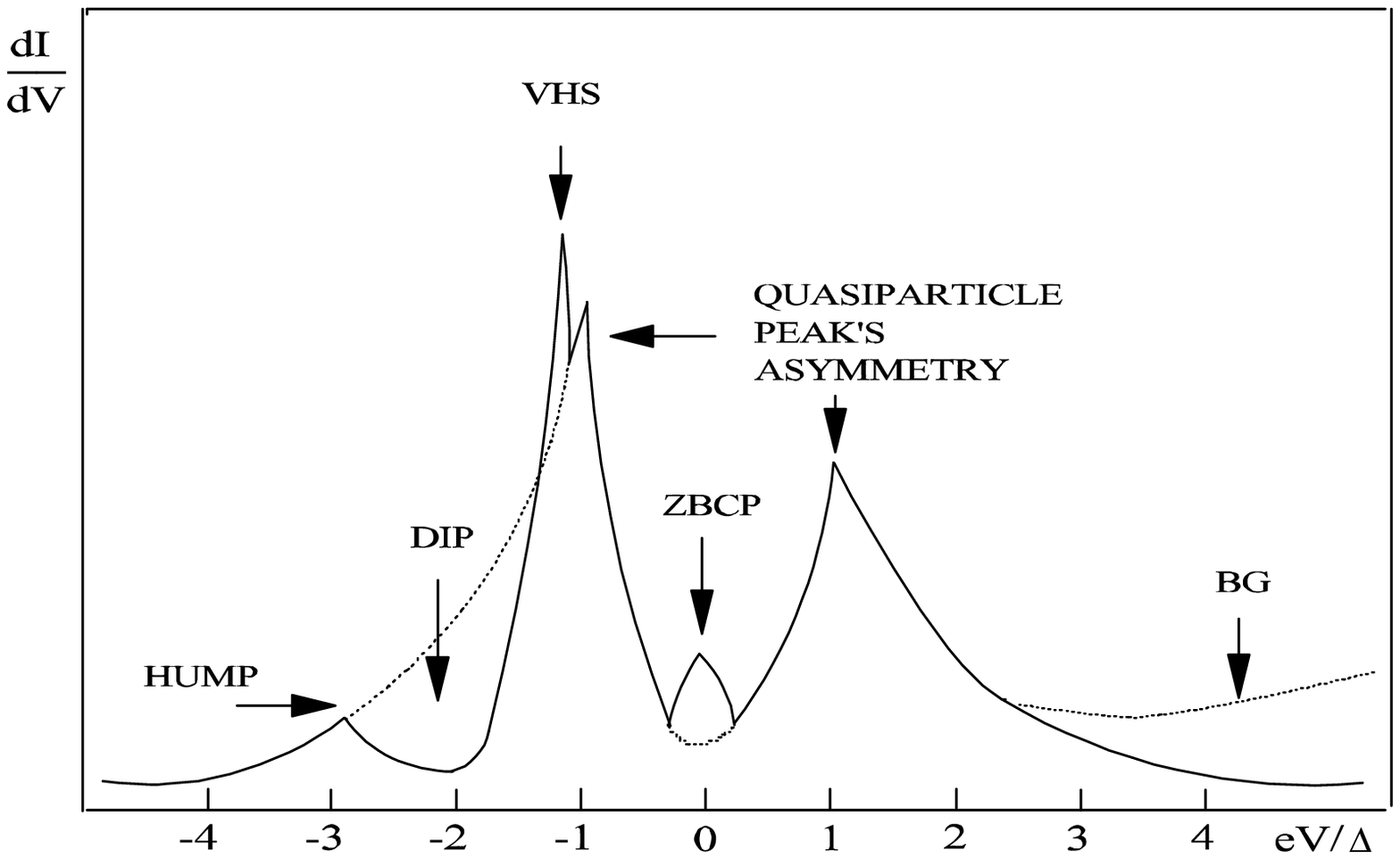}}
{\bf Fig.1} Schematic $\frac{dI}{dV}$-characteristic of NIS-structure with the main features.
\label{Fig1}
\end{figure}
While the tunneling spectroscopy on
conventional superconductors allows directly to find the energy gap of
superconductor, the same measurements in HTSC are not as easily interpreted. Some times the same
experiments on the same samples show different results [9] : cusplike or flat subgap feature, symmetric or asymmetric
conductance peaks. Usually the sharpest gap features are obtained when BG
is weakly decreasing. A quantitative measure of it is the ratio
of the conductance peak height (PH) to the background conductance:
 PHB=PH/BG.
When the BG conductance is decreasing, the  $PHB>2$, but when
BG conductance is linearly increasing ($\sim$V),
$PHB<2$ . Kouznetsov and Coffey [10] and Kirtly and Scalapino [11]
suggested that the linearly increasing BG is arising from inelastic tunneling.
As was mentioned in [1], the conductance is dominated by quasiparticle tunneling and that the effect of Andreev reflection is not significant.
A theoretical model for tunneling spectroscopy employing tight-binding
band structure, $d_{x^2-y^2}$ gap symmetry, group velocity and tunneling
directionality was studied by Z. Yusof, J. F. Zasadzinski, L. Coffey and
N. Miyahawa [1]. An angle resolved photoemission spectroscopy (ARPES)
band structure specific to optimally-doped BSCCO (Bi-2212) was used to
calculate the tunneling density of states for a direct comparison to the
experimental tunneling conductance. This model produces an asymmetric,
decreasing conductance background, asymmetric conductance peaks and
variable subgap shape, ranging from sharp, cusplike to a flat, BCS-like
feature.
A standard technique in analyzing the tunneling conductance
is to use a smeared BCS function
\begin{equation}
N(E)=N(0)\frac{E-i\Gamma}{\sqrt{(E-i\Gamma)^2-\Delta^2}}
\end{equation}
in which a scattering rate parameter (lifetime broadening factor) $\Gamma$ is used to take into account
any broadening of the gap region in the DOS. Fig. 2 shows the DOS
calculated by formula (1) at $\Delta=46 $ meV and $\Gamma=9$ meV (a),
$\Gamma=3$ meV (b) and $\Gamma=0$ (c). The characteristic features
of the DOS is the flat subgap structure at small
$\Gamma$. This method can not explain the asymmetry of the conductance peaks observed in the
tunneling experiments.

In the case of d-wave symmetry we have
\end{multicols}

\begin{figure}
\nonumber
\begin{center}
\leavevmode
\hbox{%
\epsfxsize=2.4in
\epsffile{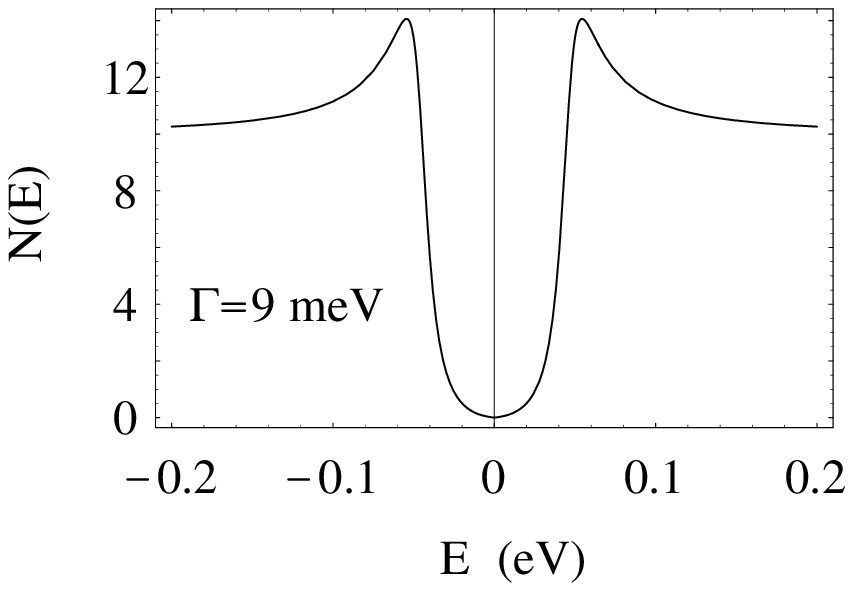}
\epsfxsize=2.3in
\epsffile{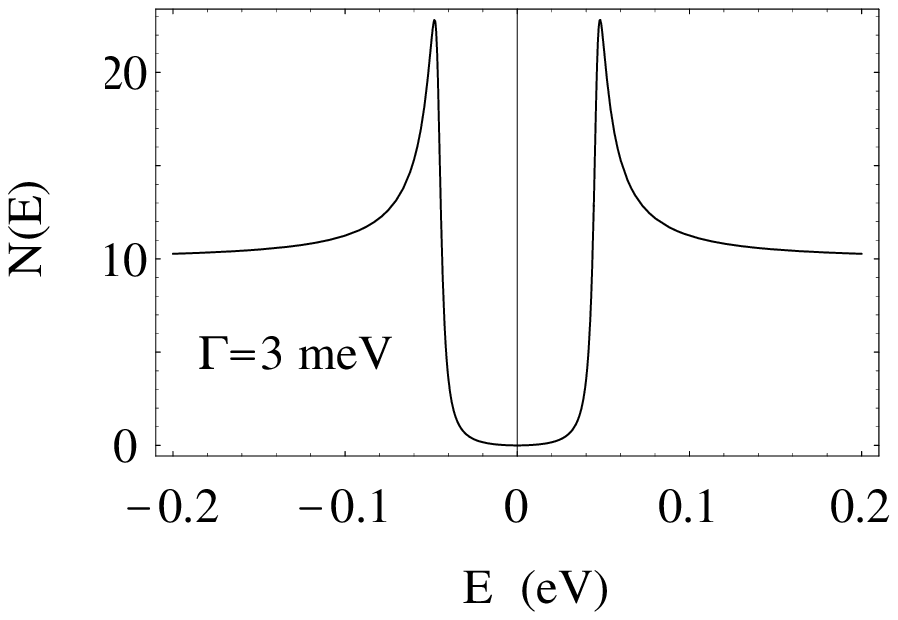}
\epsfxsize=2.3in
\epsffile{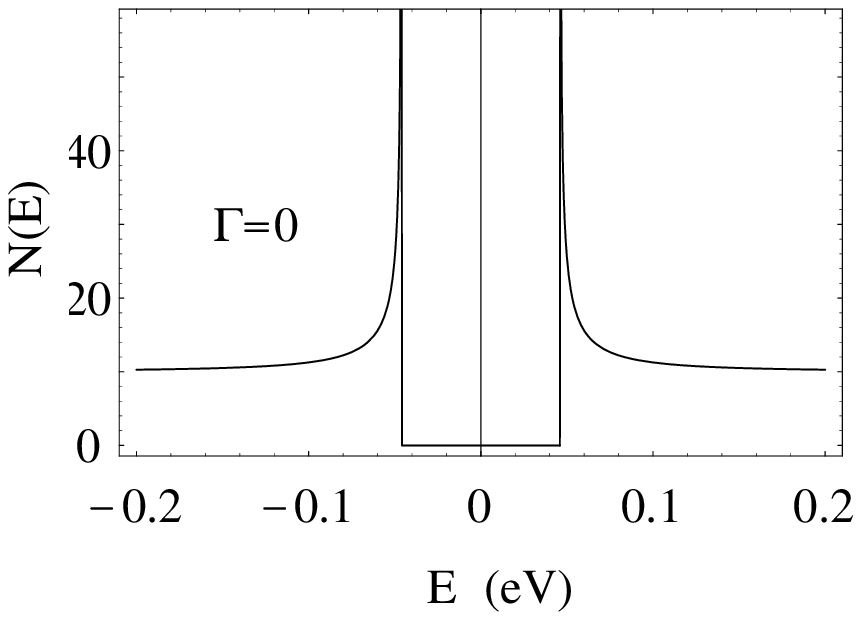}}
\end{center}
\end{figure}
\vspace{-0.8cm}
{\bf Fig.2} s-wave DOS at $\Delta=46$ meV and different values of $\Gamma$, calculated by formula (1).
\begin{figure}
\begin{center}
\leavevmode
\hbox{%
\epsfxsize=2.3in
\epsffile{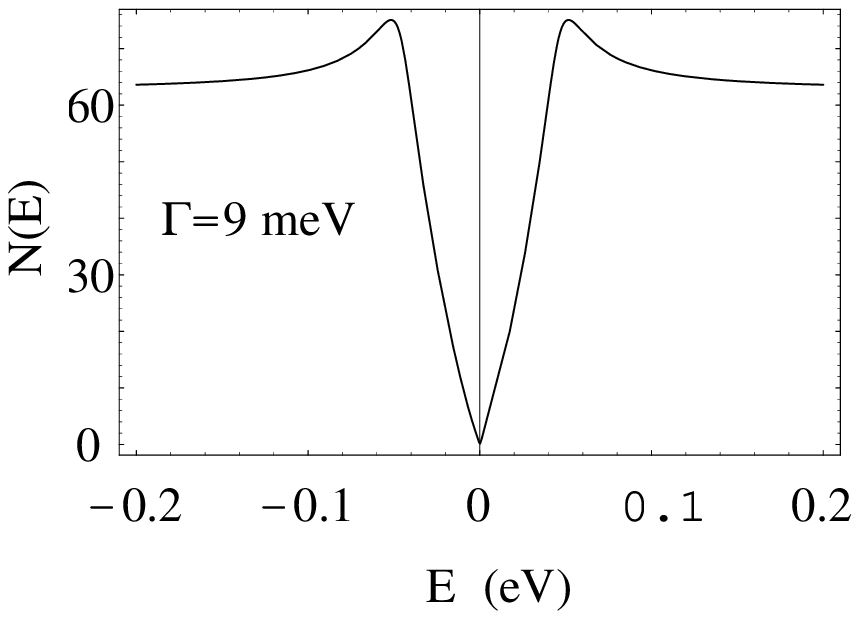}
\epsfxsize=2.3in
\epsffile{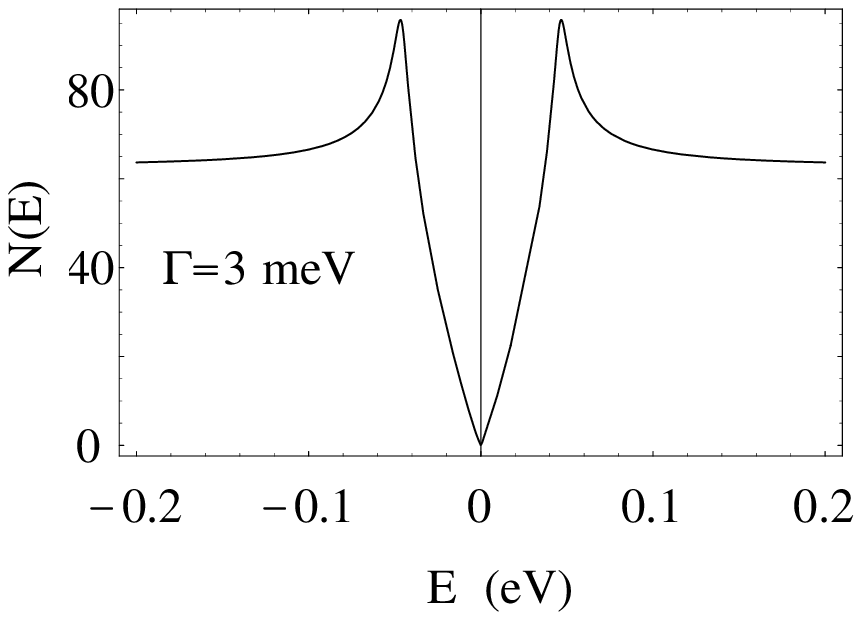}
\epsfxsize=2.4in
\epsffile{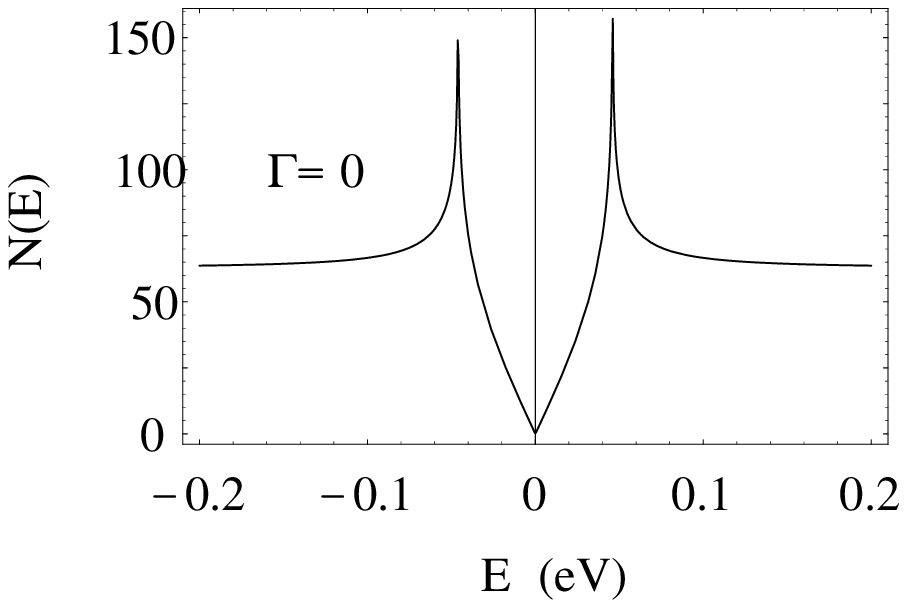}}
\end{center}
\end{figure}
\vspace{-0.8cm}
{\bf Fig.3}  d-wave DOS at $\Delta=46$ meV and different values of $\Gamma$, calculated by formula (2).
\vspace{0.3cm}
\begin{multicols}{2}
\begin{equation}
N(E)=N(0)Re\int_{0}^{2\pi}\frac{d\phi}{2\pi}
\frac{E-i\Gamma}{\sqrt{(E-i\Gamma)^2-\Delta_{0}^2cos^{2}(2\phi)}}
\end{equation}

\noindent and DOS calculated by this formula are presented in Fig. 3.
The characteristic features of the DOS is the cusplike subgap structure. As was mentioned
in [1] this standard technique requires that the comparison be made with
normalized tunneling conductance date, and since HTSC tunneling
conductance can exhibit a varied and complex background shape, this
procedure may "filter out" too much information from the conductance
data. An alternative is to simply normalize the data by a constant.\par
In [8] the tunneling data were first normalized by constructing a
"normal state" conductance obtained by fitting the high bias data to a
third order polynomial. The normalized conductance date were compared
to a weighted momentum averaged d-wave DOS
\begin{equation}
N(E)=\int f(\phi)\frac{E-i\Gamma}
{\sqrt{(E-i\Gamma)^2-\Delta_{0}^2cos^{2}(2\phi)}}d\phi
\end{equation}
Here $f(\phi)$ is an angular  weighting function, which allows for a better fit
with the experimental date in the gap region. A weighting function
$f(\phi)=1+0.4cos(4\phi)$ was used which imposed a preferential angular
selection of the DOS along the absolute maximum of the d-wave gap and
tapers off towards the nodes of the gap. This is a rather weak directional
function since the minimum of $f(0)$ along the nodes of the d-wave
gap is still none-negligible [8].\par

A. J. Fedro and D. Koelling [12] have done the modeling of the normal state
and superconducting DOS of HTSC, using tight-binding band structure,
including the next nearest neighbors
\begin{equation}
\xi_k=-2t(cos(k_xa)+cos(k_ya))+4t'cos(k_xa)cos(k_ya)-\mu
\end{equation}

The calculation showed two singularities in DOS: a van Hove
singularity in the center of energy band due to saddle point near ($\pi$,0) at
$t'=0$ and another at the lower edge of the energy band due to extra flattening
out at (0,0).
As extended  s-wave and d-wave superconducting DOS were considered
in case of hole-doped situation ($\mu<0$) for different hole concentration.
The Fermi surface for $t'=0$ and $t'=0.45t$ at the same concentration,
corresponding $\mu/2t=-0.187$ which was used in [12] are presented in
Fig. 4a. It is needed to move up from the Fermi surface (set as the zero energy)
to reach the point ($\pi$,0) in case $t'=0$ and move down in case
$t'=0.45t$. So, for $t'=0$ the Fermi energy lies to the left of the van Hove
singularity and will move away from it with increased hole doping while for
$t'=0.45t$ it lies to the right and will move towards to with increased hole
doping (See Fig. 4b where the DOS for $t'=0$ and $t'=0.45$ are presented ).
 For calculation of the superconducting DOS Fedro and Koelling used formula
\begin{equation}
N(E)=\frac{1}{2}\sum_{k}(1+\frac{\xi_k}{E_k})\delta(E-E_k)+
(1-\frac{\xi_k}{E_k})\delta(E+E_k)
\end{equation}

This formula is the limit of the expression for tunneling
density of states (6) at $\Gamma=0$ and $|T_k|^2=1$, where $T_k$ is tunneling matrix element.
\end{multicols}
\begin{figure}
\begin{center}
\leavevmode
\hbox{%
\epsfxsize=3.0in
\epsffile{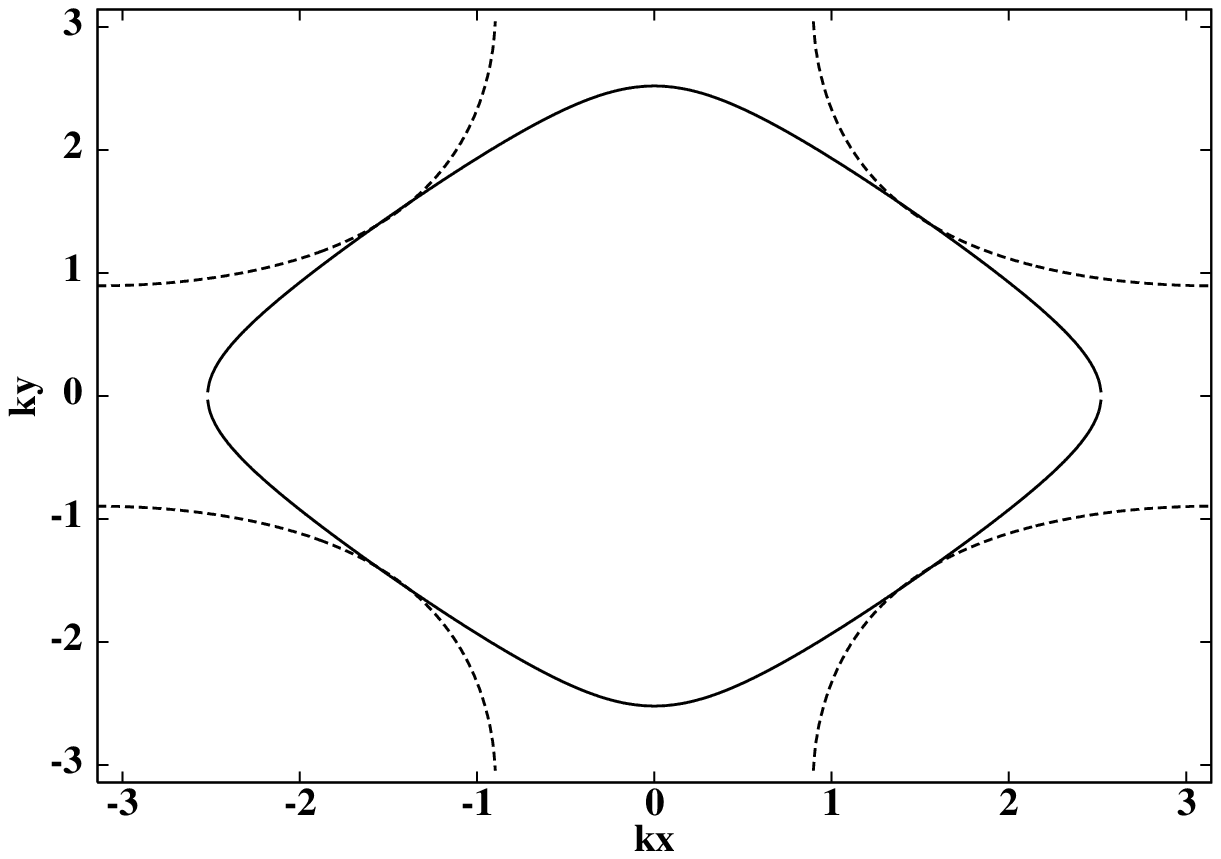}
\hspace{0.5cm}
\epsfxsize=3.0in
\epsffile{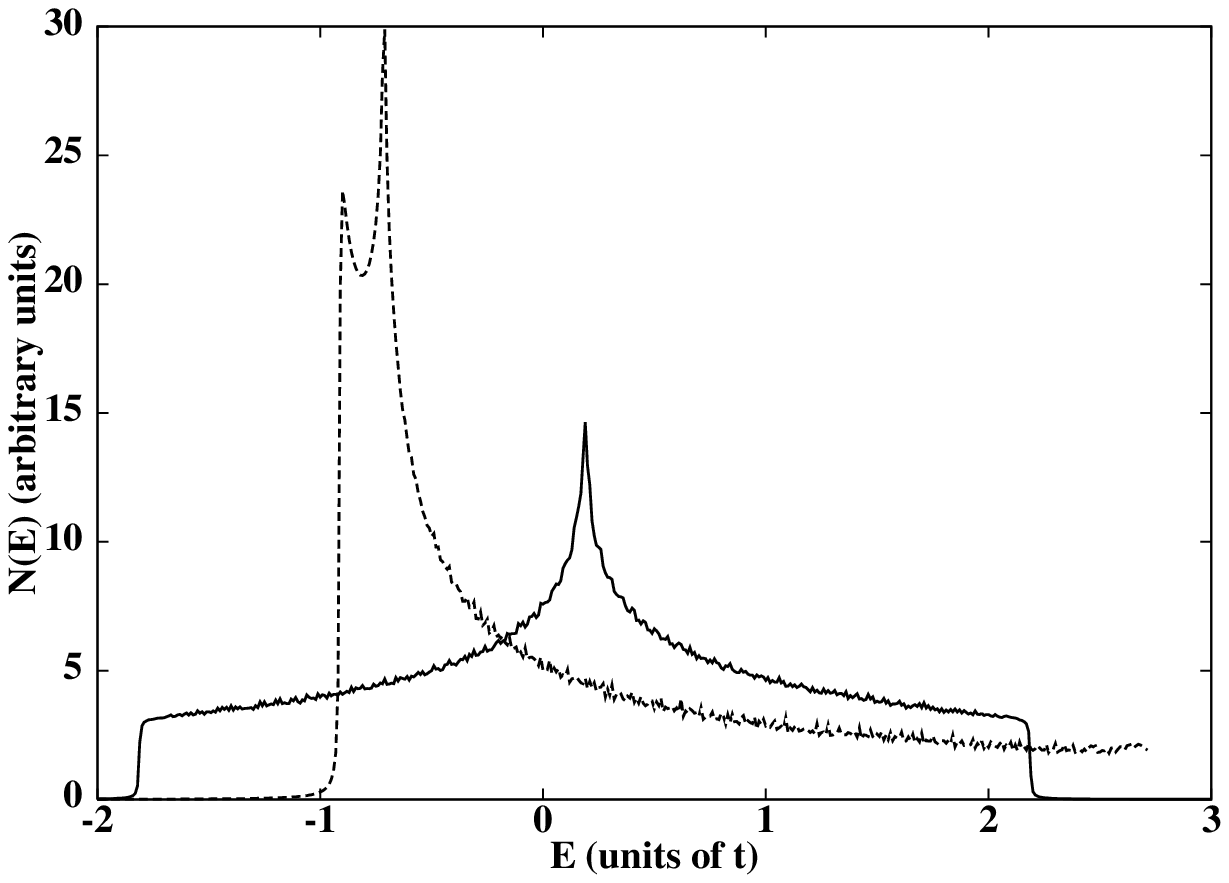}}
\end{center}
\end{figure}
\vspace{-0.6cm}
{\bf Fig.4} Fermi surfaces (left) and DOS (right) for $t'$=0 (solid lines) and $t'$=0.45t (dashed lines) in formula (4) at $\frac{\mu}{2t}$=-0.187 which corresponds
hole - doped situation.
\begin{figure}
\begin{center}
\leavevmode
\hbox{%
\epsfxsize=3.0in
\epsffile{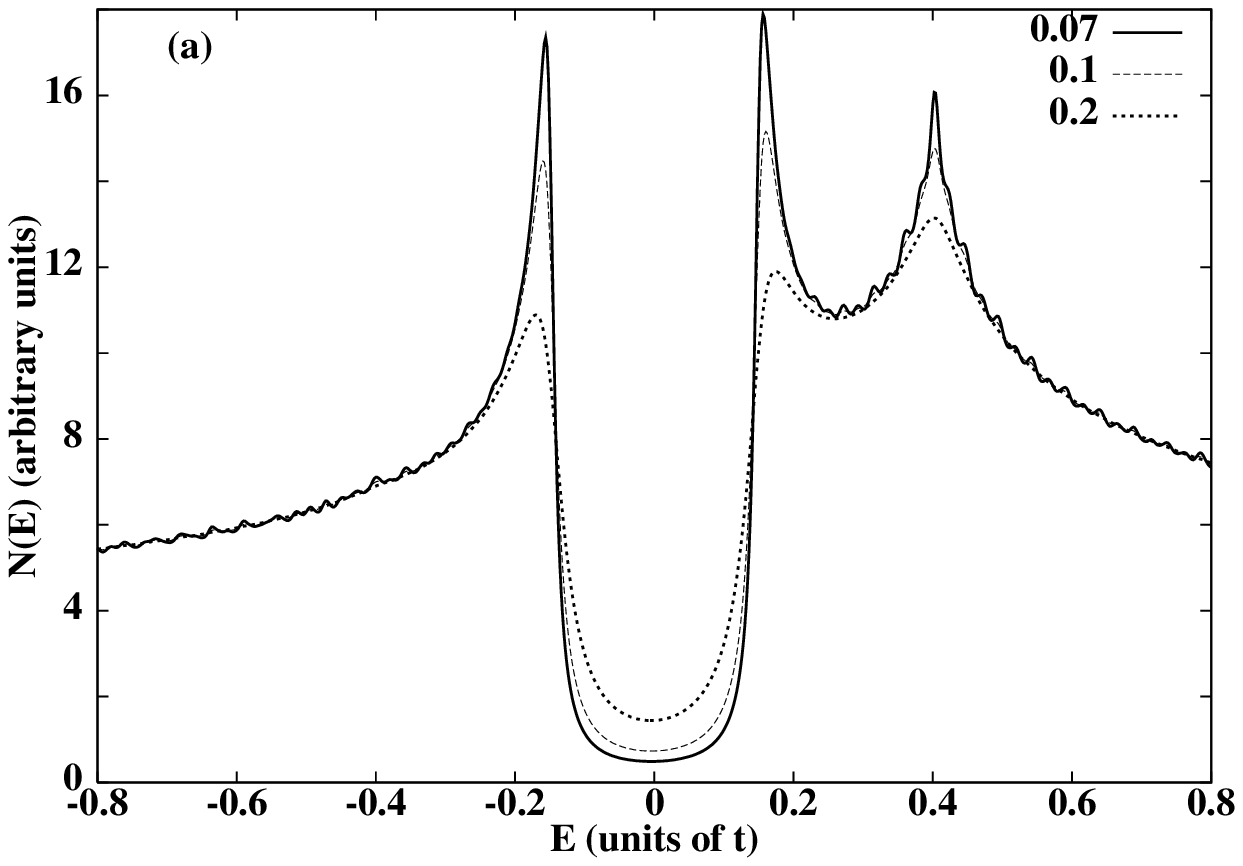}
\hspace{0.3cm}
\epsfxsize=3.0in
\epsffile{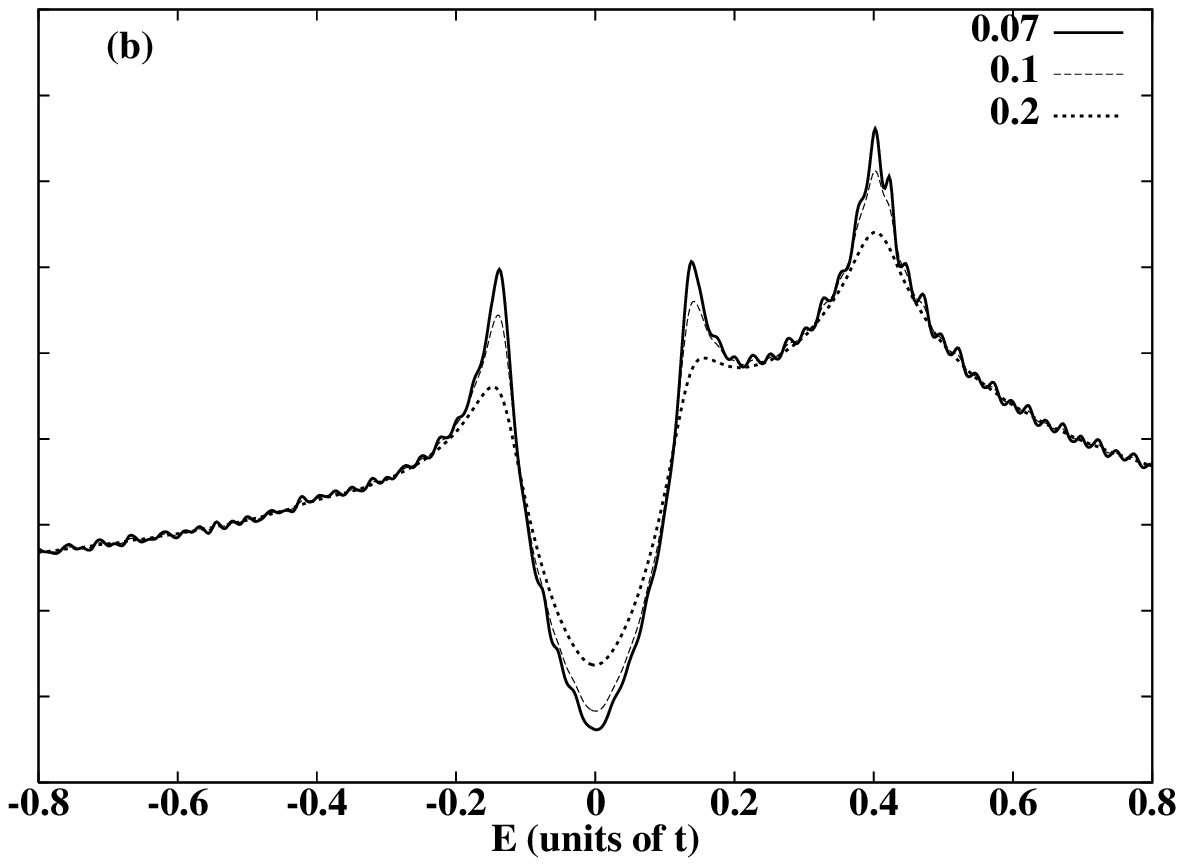}}
\end{center}
\end{figure}
\vspace{-0.6cm}
{\bf Fig.5} DOS for $t'$=0 at different $\Gamma$ for s-wave symmetry (a) and
d-wave symmetry (b), calculated by formula (6).
\begin{multicols}{2}
The Fig.5a shows the result of calculation
 of the DOS for $t'$ = 0 at $\Gamma=$ 0.07, 0.1 and 0.2 meV for s-wave
 symmetry which reflect the results of Fedro and Koelling. Fig.5b shows the same DOS for d-wave symmetry. In both cases the Fermi energy
(set as the zero of energy) lies to the left of the van Hove singularity.
There is the peaks' asymmetry which is more pronounced at large $\Gamma$.
\section{  Models and Methods}
In this paper we use the method for calculation of the DOS presented in [1].
The tunneling DOS of a superconductor is determined by the imaginary part
of the retarded single particle Green's function
\begin{equation}
N(E)=-\frac{1}{\pi}Im\sum_{k}|T_k|^2G^{R}(k,E)
\end{equation}
For the superconducting state
\begin{equation}
G^{R}(k,E)=\frac{u_{k}^{2}}{E-E_k+i\Gamma}+
\frac{v_{k}^{2}}{E+E_k+i\Gamma}
\end{equation}
where $u_k^2$ and $v_k^2$ are the usual coherence factors,
\begin{eqnarray}
u_k^2=\frac{1}{2}(1+\frac{\xi_k}{E_k})\cr
v_k^2=\frac{1}{2}(1-\frac{\xi_k}{E_k})
\end{eqnarray}
and $\Gamma$ is the
quasiparticle lifetime broadening factor. The energy spectrum of
quasiparticles in the superconducting state is determined by
\begin{equation}
E_k=\sqrt{|\Delta(k)|^2+\xi_k^2}
\end{equation}
with the effective band structure extracted from ARPES experiments [13]
\begin{eqnarray}
\xi_k=C_0+0.5C_1[cos(k_xa)+cos(k_ya)]~~~~~~~~~~~~~~~~~~~~\cr
~~+C_2cos(k_xa)cos(k_ya)+
0.5C_3[cos(2k_xa)+cos(2k_ya)]\cr
~+0.5C_4
[cos(2k_xa)cos(k_ya)+cos(k_xa)cos(2k_ya)]\cr
+C_5cos(2k_xa)cos(2k_ya)
\end{eqnarray}
Here $\xi_k$ is measured with respect to the Fermi energy ($\xi_k$=0), and
the phenomenological parameters are (in units of eV) $C_0=0.1305$,
$C_1=-0.5951$, $C_2=0.1636$, $C_3=-0.0519$, $C_4=-0.1117$, $C_5=0.0510$.\par
\end{multicols}
\begin{figure}
\begin{center}
\leavevmode
\hbox{%
\epsfxsize=2.1in
\epsffile{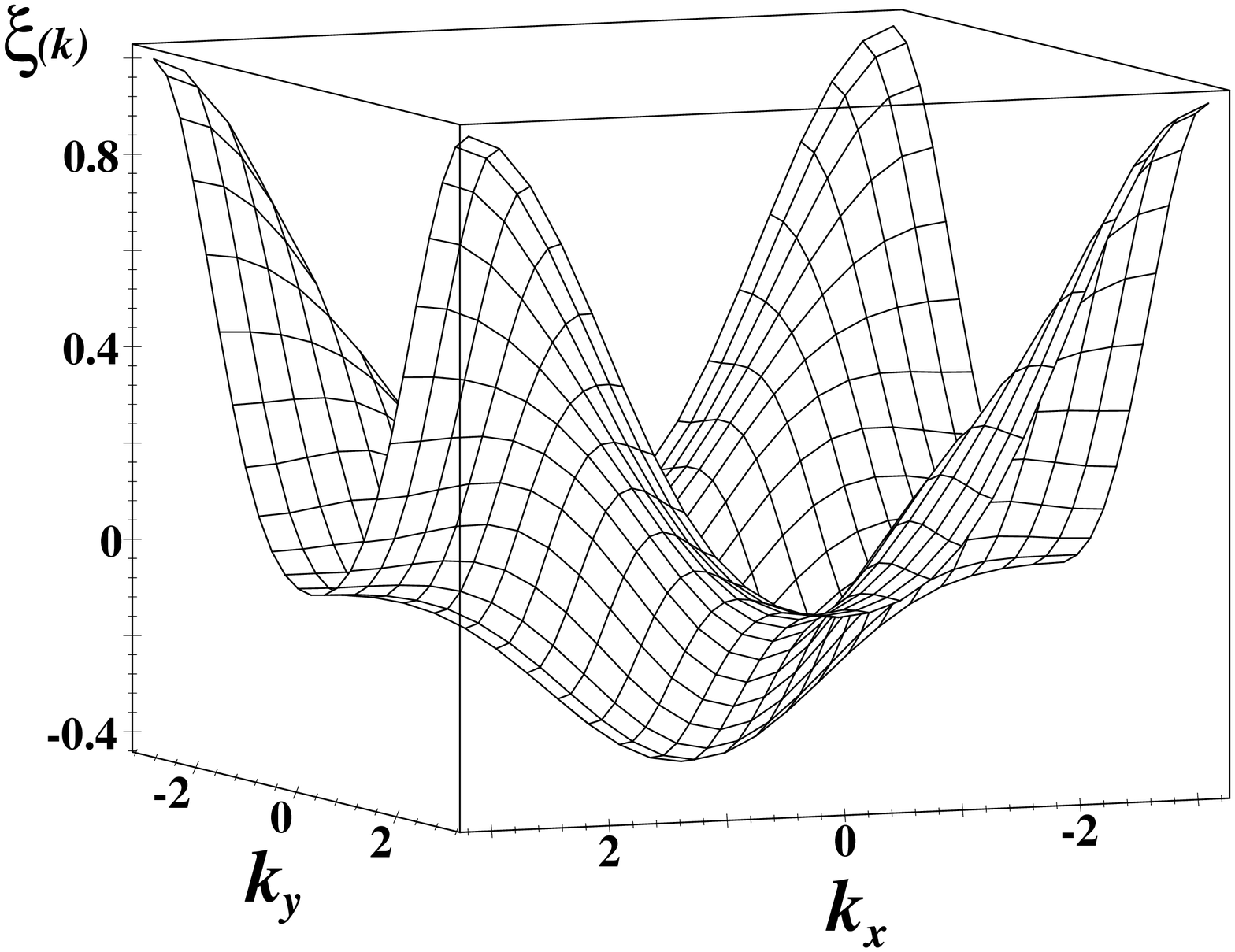}
\hspace{0.5cm}
\epsfxsize=2.1in
\epsffile{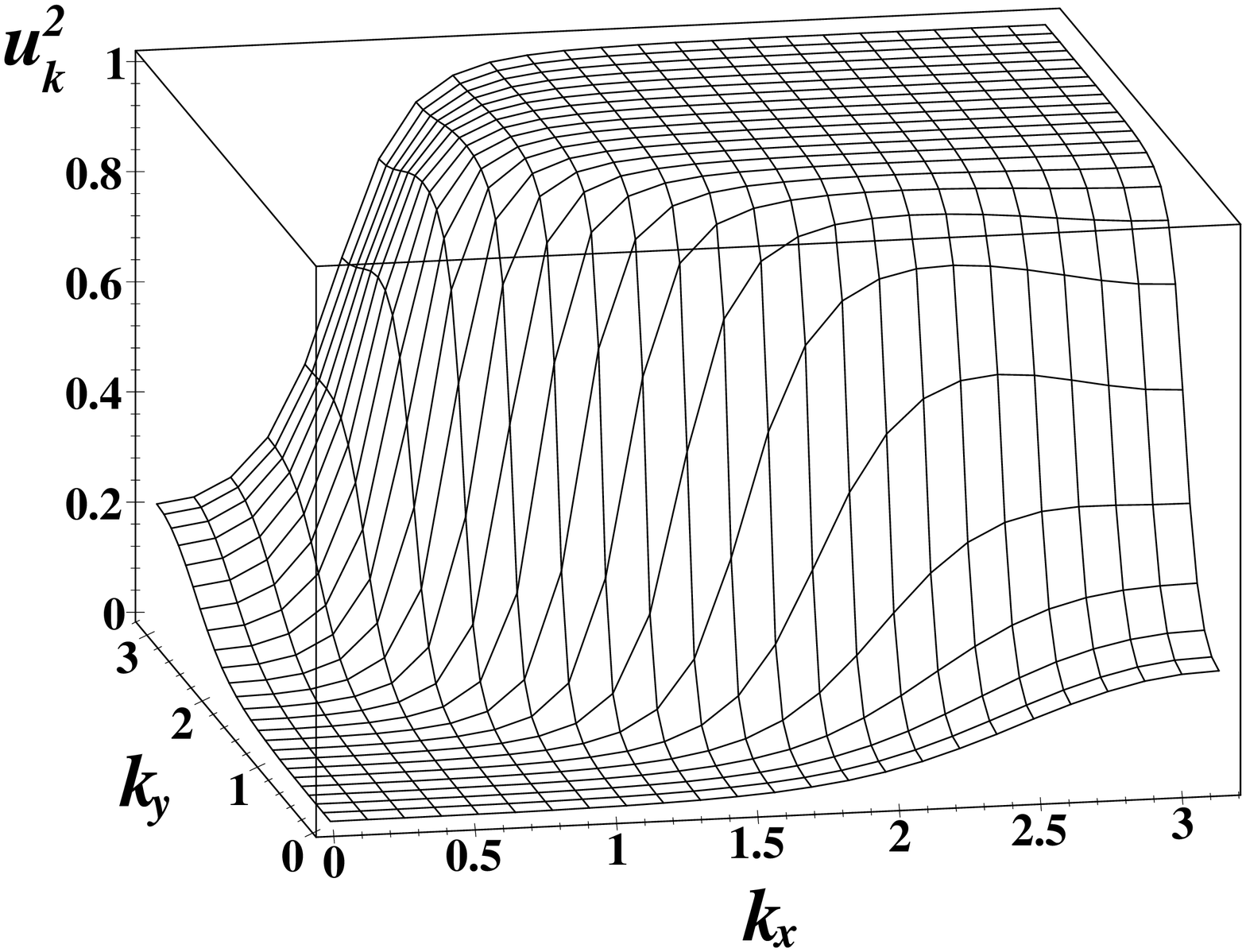}
\hspace{0.5cm}
\epsfxsize=1.6in
\epsffile{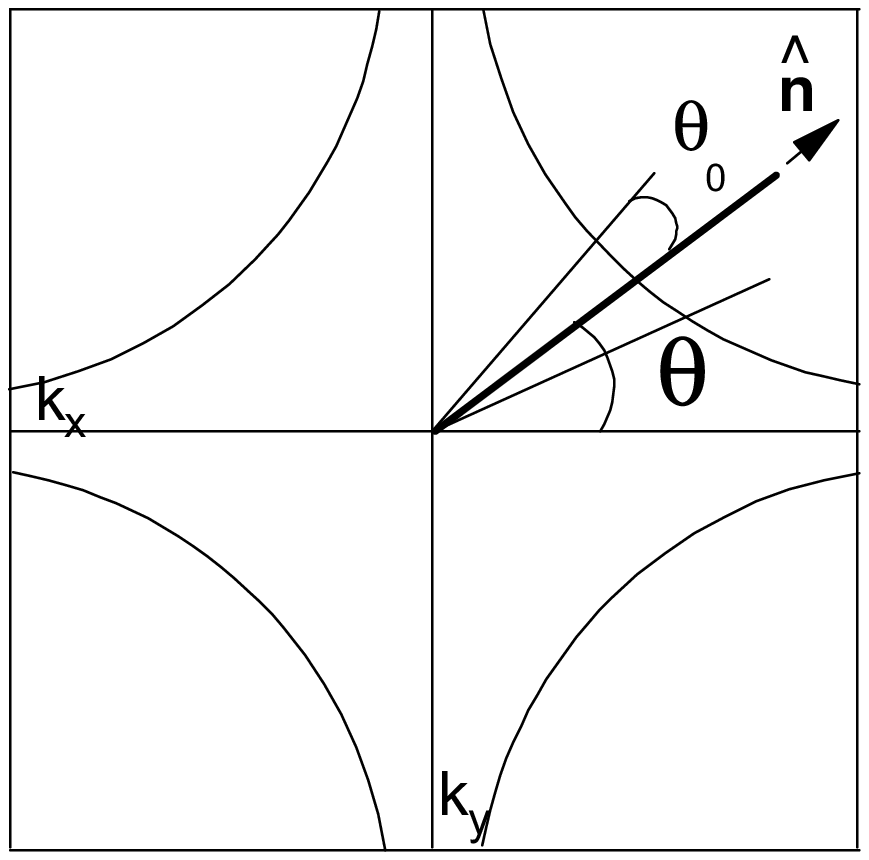}}
\end{center}
\end{figure}
\vspace{-0.6cm}
\noindent {\bf Fig.6}(left) 3D-plot of energy spectrum of normal
state according to formula (9).\\
{\bf Fig.7}(center) 3D-plot of coherence factor $u_k^2$ according to formula (8).\\
{\bf Fig.8}(right) Fermi surface corresponding to the  $\xi_k=0$ in formula (10). Dark straight line shows the line of directional tunneling, the dashed lines show the angular spread $\Theta_0$.
\begin{multicols}{2}
Fig. 6 shows the three dimensional image of function (10) .
There are saddle
point in ($\pi$,0) and flattening out of the energy band at (0,0) which
lead to the van Hove singularities in the DOS. The three dimensional graph of the coherence factor $u_k^2$
is shown in Fig. 7.\par
Since quasiparticles with momentum perpendicular to the barrier interface
have the highest probability of tunneling, the tunneling matrix element $|T_k|^2$ reveals
a need for factors of directionality $D(k)$ and group velocity $v_g(k)$ [1]. The group velocity
factor is defined by
\begin{equation}
v_g(k)=|\vec \nabla_k\xi_k.\hat n|=
|\frac{\partial\xi_k}{\partial k_x}cos(\theta)+
\frac{\partial\xi_k}{\partial k_y}sin(\theta)|
\end{equation}
where the unit vector $n$ defines the tunneling direction as  shown in Fig. 8,
which is perpendicular to the plane of the junction.\par
The directionality function $D(k)$ is defined by
\begin{equation}
D(k)=exp[-\frac{k^2-(\bf k.\hat n)^2}{( {\bf k.\hat n})^2\Theta_0^2}]
\end{equation}
Here $\Theta_0$ defines the angular spread of the quasiparticle momentum with none-negligible tunneling probability
with respect to n. The tunneling matrix element $|T_k|^2$ is written as
\begin{equation}
|T_k|^2=v_g(k)D(k)
\end{equation}
The three dimensional graphs of group velocity $v_g(k)$, directionality $D(k)$
and tunneling matrix element $|T_k|^2$
functions are shown in Fig 9.
\section{Results and discussions}
Different factors may lead to the changing of the energy gap $\Delta _0$ in HTSC.
In particular, strong effects are caused by nonmagnetic impurities [14].
In superconductors with d-wave symmetry the nonmagnetic impurities
destroy the superconductivity very efficiently. Possibility to destroy of Cooper pairs by
impurities leads to their finite lifetime.
If the state with the quasiparticle is not stationary state, it must attenuate with time
due to transitions to other states.
The corresponding wave function has the form
$e^{-i\xi (p)t/\hbar-\Gamma t/\hbar}$, where $\Gamma$ is
proportional to the probability of the transitions to the other states.
It may be interpreted as the energy of the quasiparticle
has the imaginary addition $-i\Gamma$. The relation between $\Gamma$ and
\end{multicols}
\begin{figure}
\begin{center}
\leavevmode
\hbox{%
\epsfxsize=2.1in
\epsffile{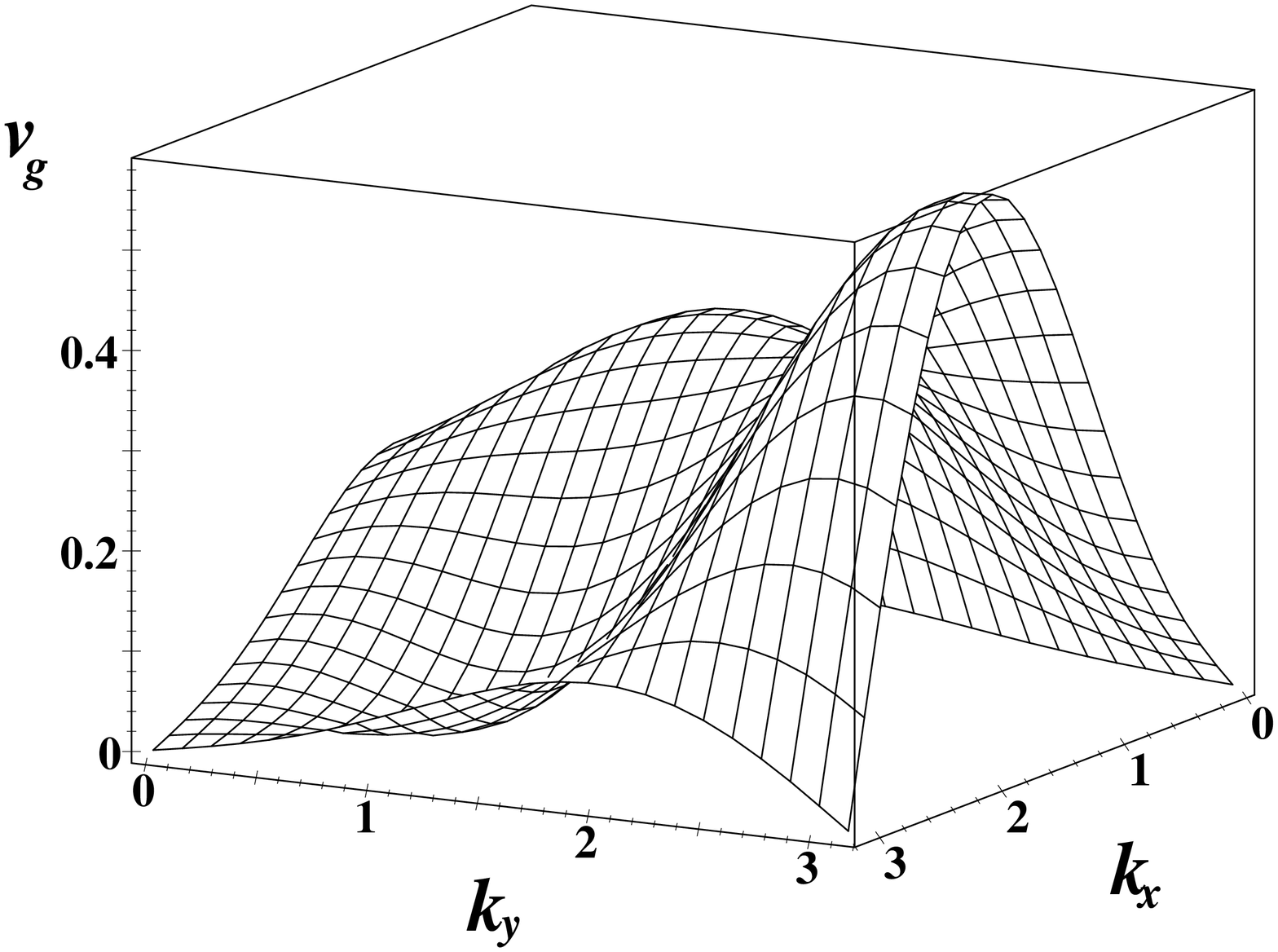}
\epsfxsize=2.3in
\epsffile{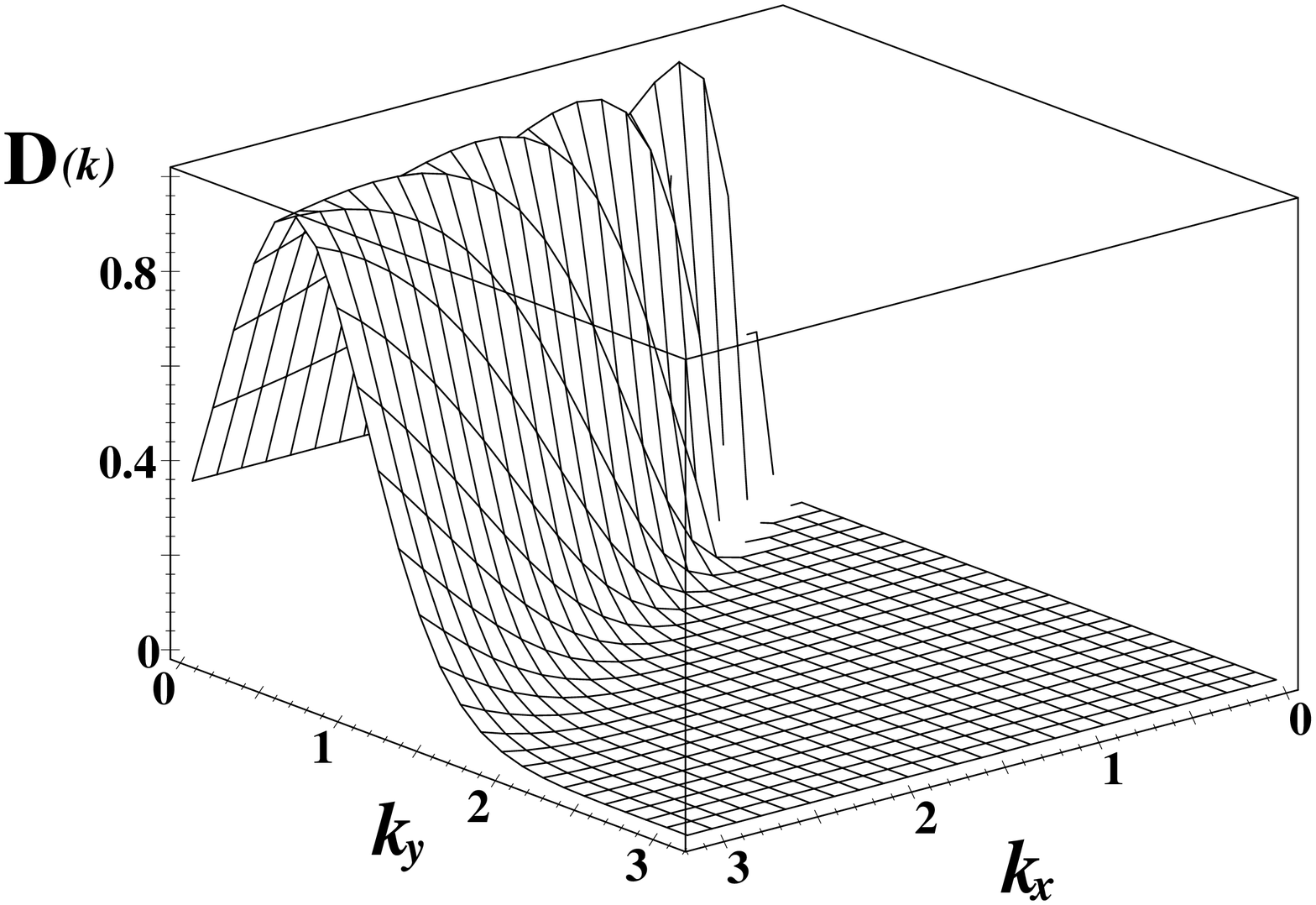}
\epsfxsize=2.0in
\epsffile{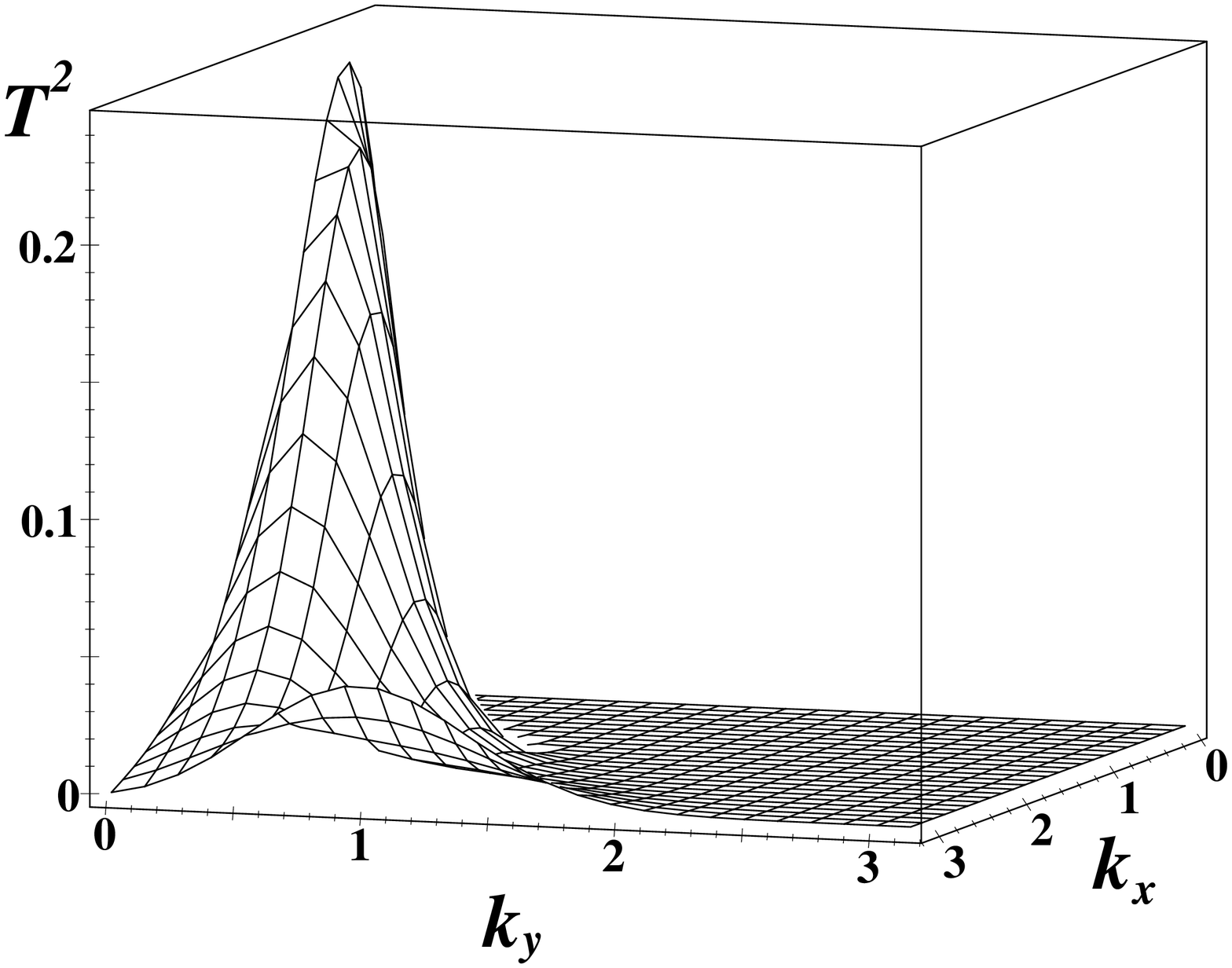}}
\end{center}
\end{figure}
\vspace{-0.6cm}
{\bf Fig.9} 3D plot of the group velocity function (a), the directionality function
(b) and the tunneling matrix element according to formulas (11), (12) and (13),
correspondingly.
\begin{figure}
\begin{center}
\leavevmode
\hbox{%
\epsfxsize=3.0in
\epsffile{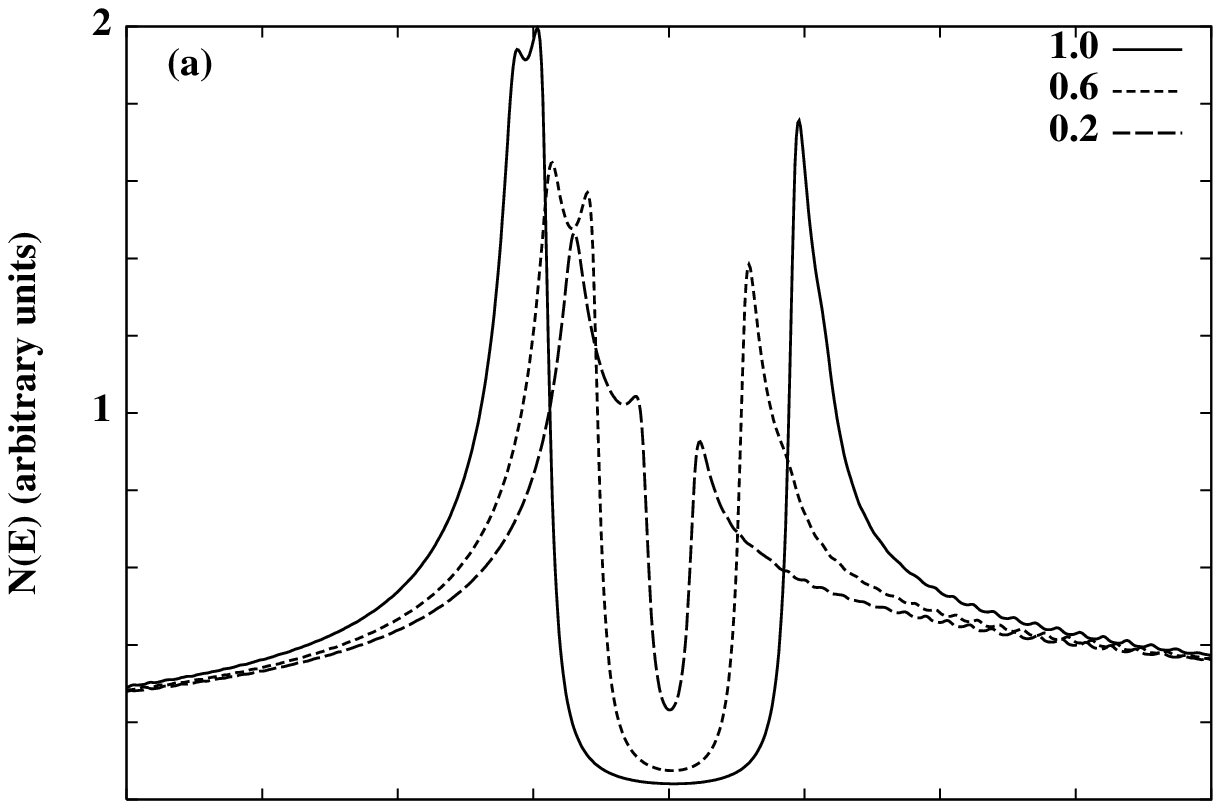}
\hspace{-0.6cm}
\epsfxsize=3.0in
\epsffile{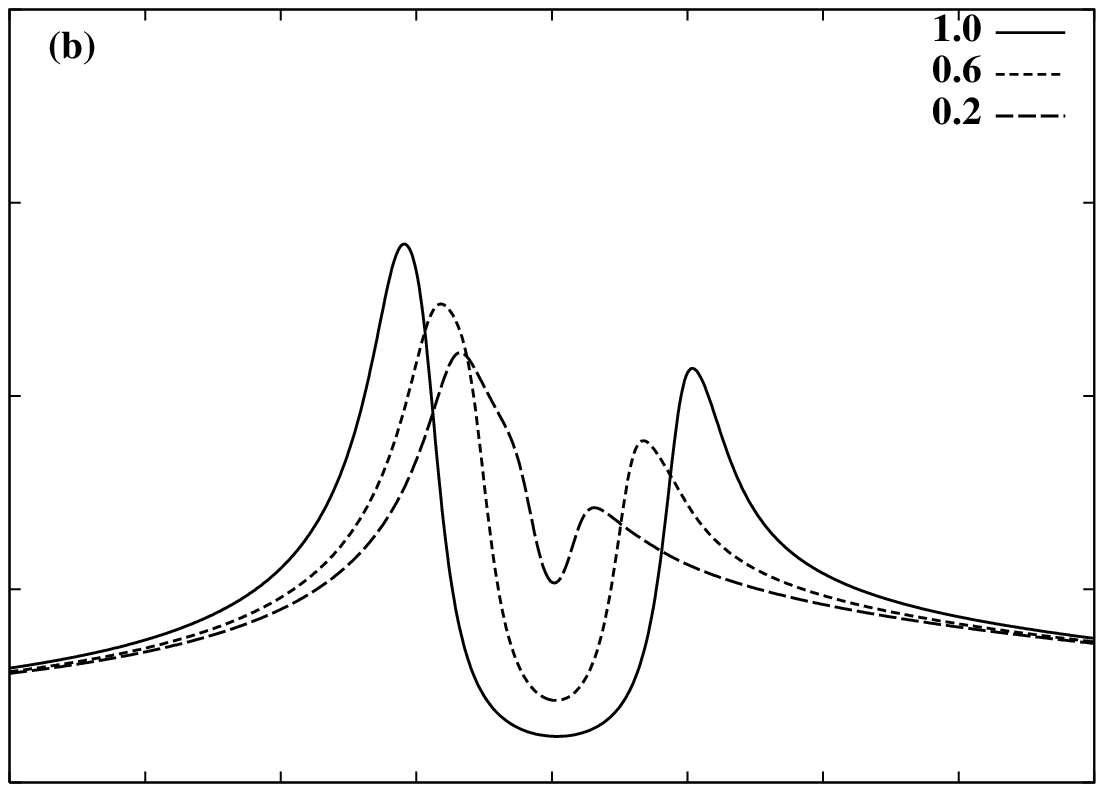}}
\end{center}
\end{figure}
\vspace{-1.78cm}
\begin{figure}
\begin{center}
\leavevmode
\hbox{%
\epsfxsize=3.0in
\epsffile{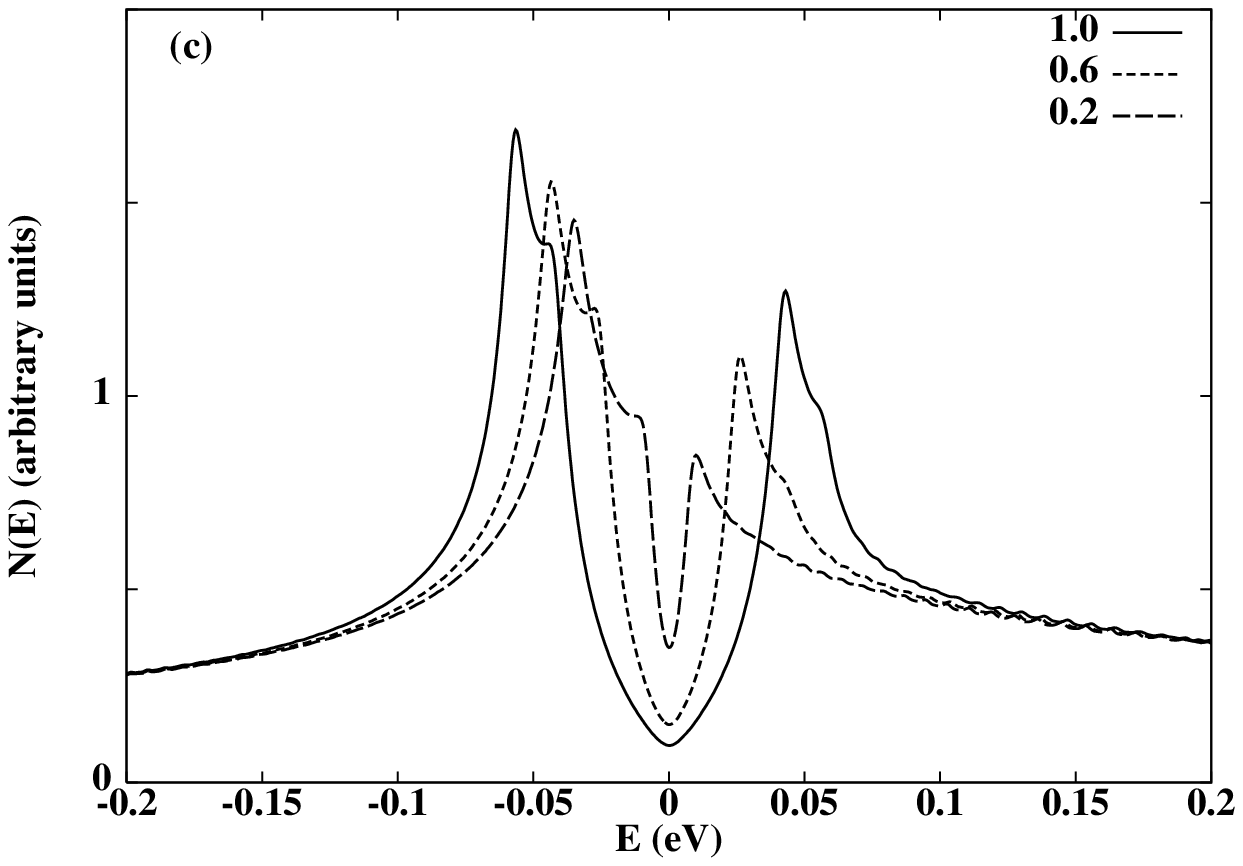}
\hspace{-0.6cm}
\epsfxsize=3.0in
\epsffile{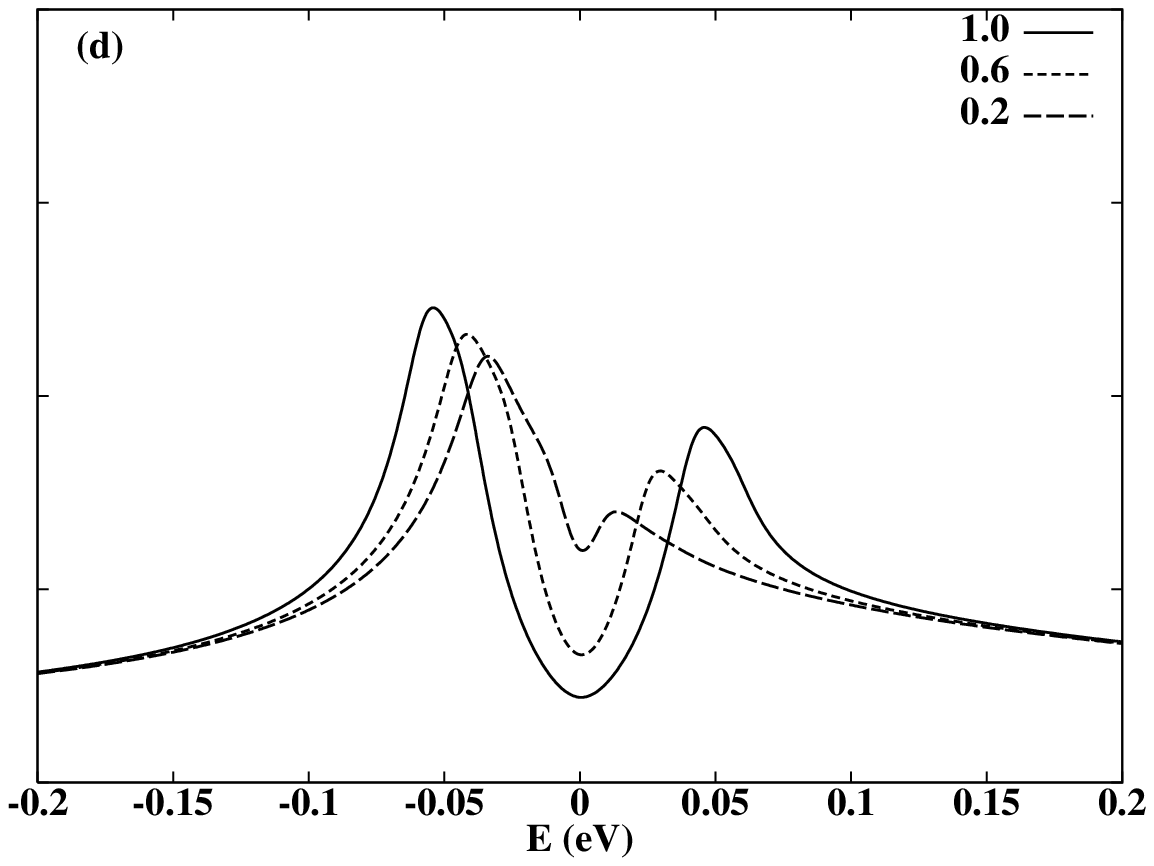}}
\end{center}
\end{figure}
\vspace{-0.6cm}
{\bf Fig.10} The changing of DOS with energy gap $\Delta_0$ for s- (a,b) and d-wave (c,d) symmetry at  $\Gamma$=3 meV (a,c) and $\Gamma$=9 meV (b,d) without of effects of directionality and group velocity.
\begin{multicols}{2}
\noindent lifetime of quasiparticle  $\tau_s$
is $\Gamma=\hbar/\tau_s$. Hence, the impurities lead to changing $\Delta_0$ and
we may do modeling of the
influence of impurities on tunneling conductance by numerical calculation of DOS
N(E) considering different
values of $\Delta_0$ in formula (6). Here we present results of calculation N(E) at
$\Delta_0=\alpha\Delta_{00}$,
where $\alpha$=0.2, 0.4, 0.6, 0.8, 1 and $\Delta_{00}$= 46 meV.\par
The peculiarities of the quasiparticle energy spectrum (10) play an essential role in
explanation of the
conductance features. Here, based on the numerical calculation of DOS
we consider that
the underlying asymmetry of the conductance peaks is primarily due to the features
of quasiparticles energy
spectrum. The d-wave gap symmetry simply enhances the degree of the peaks
asymmetry. The last one is also
changed by changing the tunneling direction.\par
Fig. 10 shows the results of the numerical calculations of the DOS at
$\Gamma _0 =3$ meV (a,c) and $\Gamma _0 =9$ meV (b,d) without effects
of group velocity and directionality as for s-wave (a,b) and d-wave (c,d) gap
symmetry, correspondingly for different values of energy gap $\Delta_0$.
We have decreased the energy gap $\Delta _0$, starting from $\Delta _0=46$meV.
For clarity we present only three characteristic curves, which corresponds to
$\alpha \Delta _0$ with $\alpha=$1, 0.6 and 0.2.
We exclude the effects of group velocity and directionality to demonstrate that
they are not responsible
for peaks asymmetry.\par
There is the asymmetry of the quasiparticle peak heights as for s- and d-wave
symmetry. So, the origin of
the peaks asymmetry is not due to d-wave symmetry of the energy gap of HTSC.
There is more flat subgap behavior of DOS in the case of s-wave symmetry in
comparing with the d-wave case. The increase of lifetime broadening factor
$\Gamma$ leads to the enhance of the peaks' asymmetry. There are van Hove
singularities in the DOS at small $\Gamma$. The increase of $\Gamma$ leads to
the confluence
of the quasiparticle and VHS peaks and this results to the enhance
of the DOS peaks asymmetry due to saddle point in energy spectrum (10)
at ($\pi$,0).
Also note to the asymmetry of the background as for s- and d-wave
gap symmetry.\par
Fig. 11 shows the $\Delta _0$-dependence of DOS taking into account the effects
of group velocity and directionality at $\Gamma=3$ meV (a,c) and
$\Gamma=9$ meV (b,d)  for s-wave (a,b) and d-wave (c,d) gap symmetry.
As in [1] we have taken $\Theta =0.25$ and $\Theta _0 =0.1$.\par
There is also quasiparticle peaks' asymmetry similar to s- and d-wave cases.
But in d-wave case the asymmetry is more stronger than in s-wave.
The effects of group velocity and directionality lead to disappear of the VHS
in DOS.
The increase of $\Gamma$ enhances the quasiparticle peak asymmetry.
The most strong effect of energy band structure on the DOS occurs along
$k_x$-axis due to van Hove singularity at $(\pi,0)$.\par
\end{multicols}
\begin{figure}
\begin{center}
\leavevmode
\hbox{%
\epsfxsize=3.0in
\epsffile{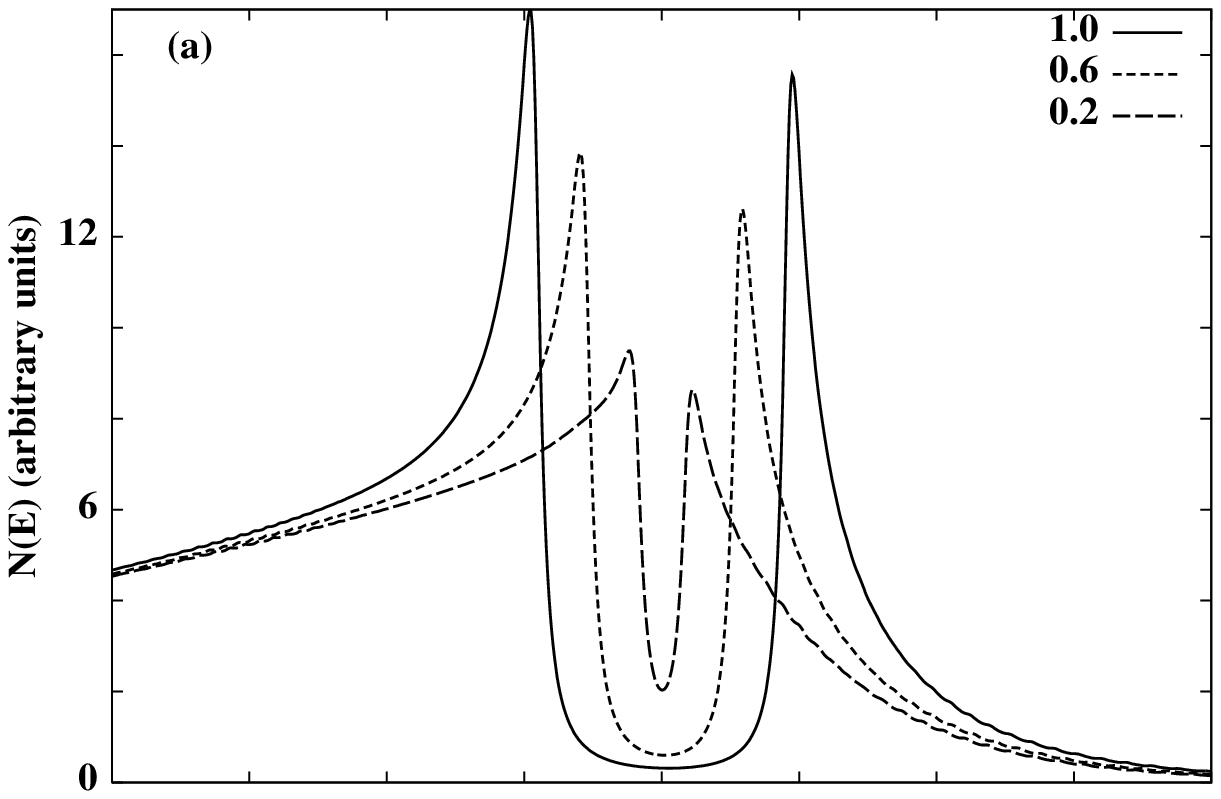}
\hspace{-0.6cm}
\epsfxsize=3.0in
\epsffile{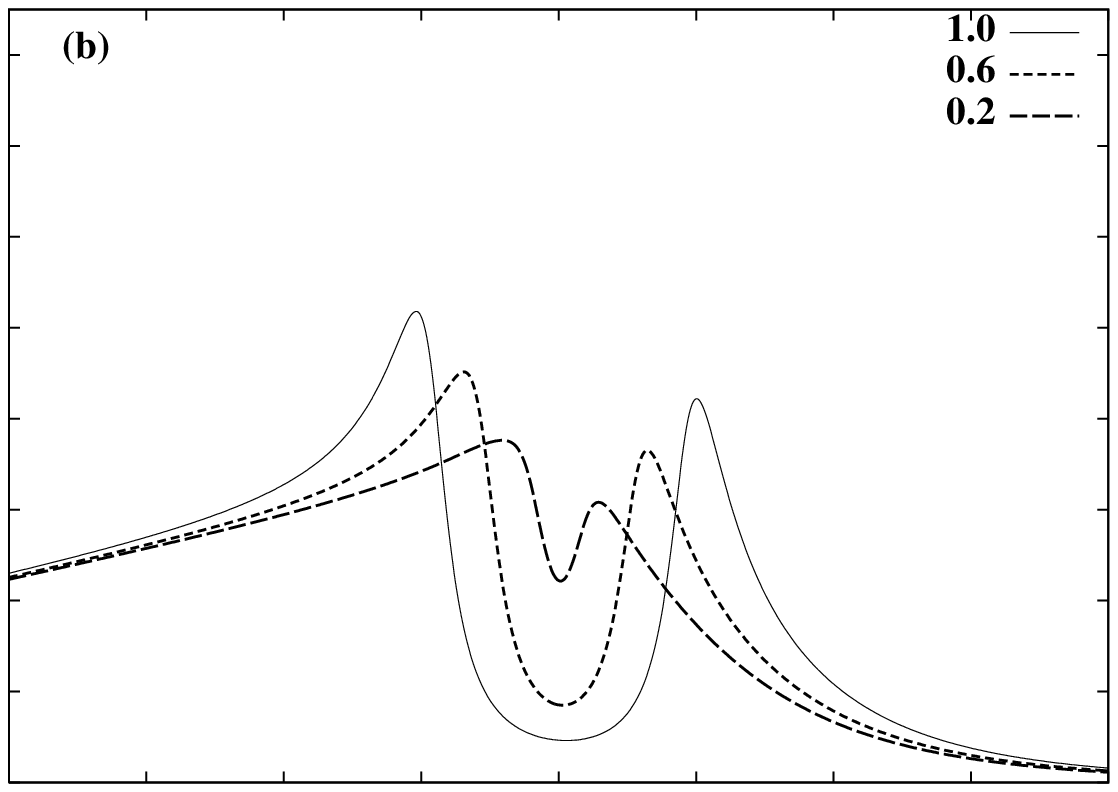}}
\end{center}
\end{figure}
\vspace{-1.8cm}
\begin{figure}
\begin{center}
\leavevmode
\hbox{%
\epsfxsize=3.0in
\epsffile{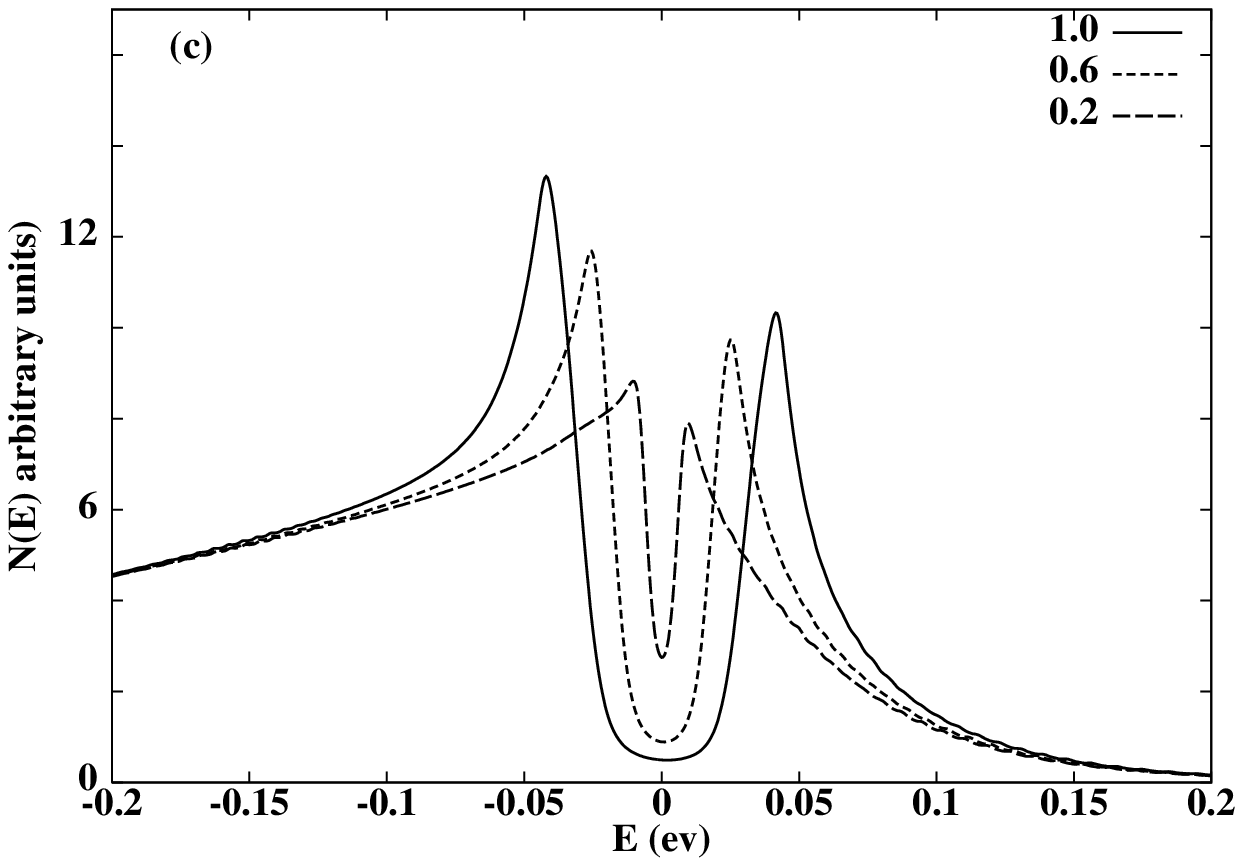}
\hspace{-0.6cm}
\epsfxsize=3.0in
\epsffile{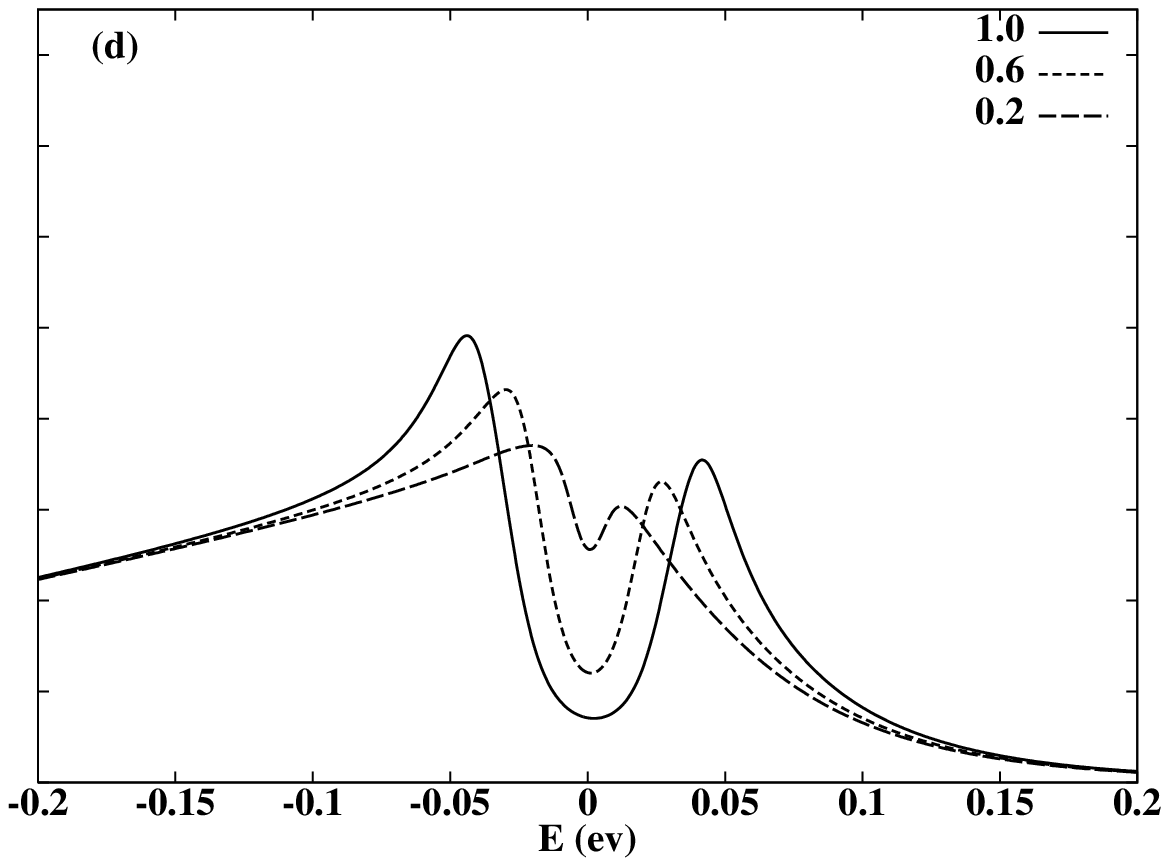}}
\end{center}
\end{figure}
\vspace{-0.6cm}
{\bf Fig.11} The changing of DOS with energy gap $\Delta_0$ for s- wave
(a,b) and d-wave (c,d) gap symmetry at  $\Gamma$=3 meV (a,c) and
$\Gamma$=9 meV (b,d) with the
 effects of directionality and group velocity.

\begin{figure}
\begin{center}
\leavevmode
\hbox{%
\epsfxsize=3.0in
\epsffile{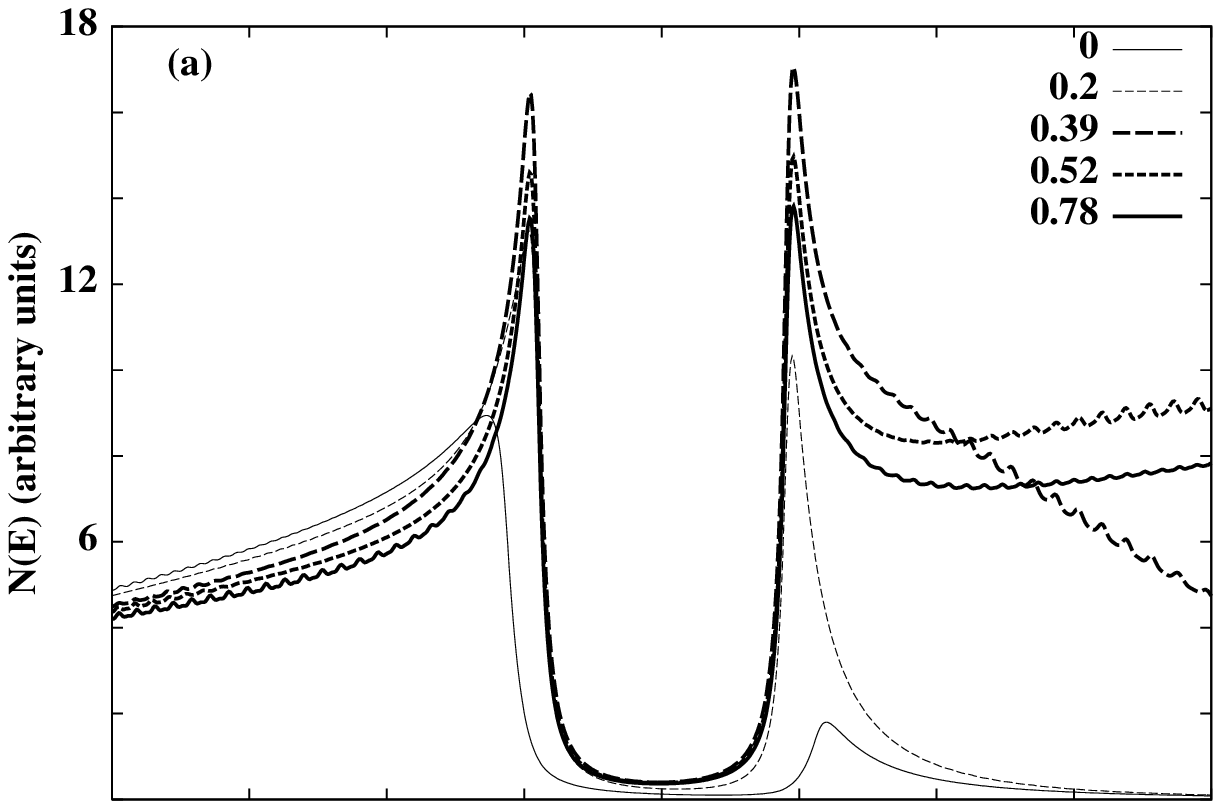}
\hspace{-0.6cm}
\epsfxsize=3.0in
\epsffile{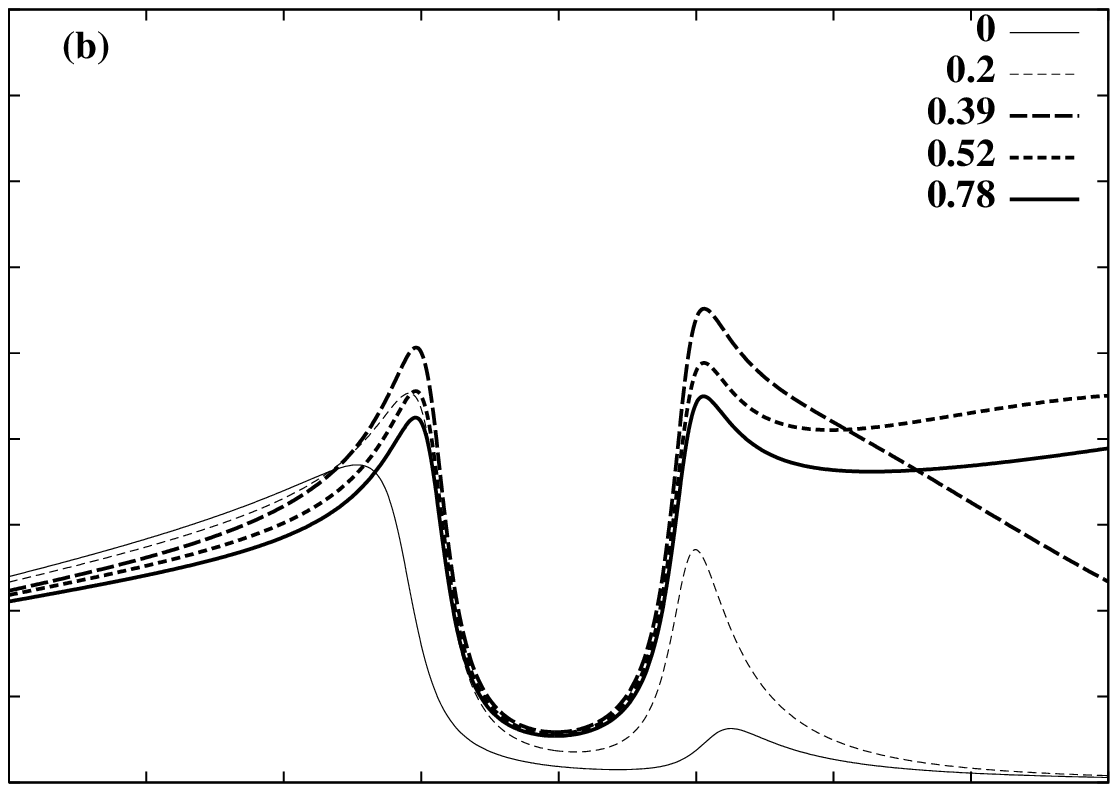}}
\end{center}
\end{figure}
\vspace{-1.78cm}
\begin{figure}
\begin{center}
\leavevmode
\hbox{%
\epsfxsize=3.0in
\epsffile{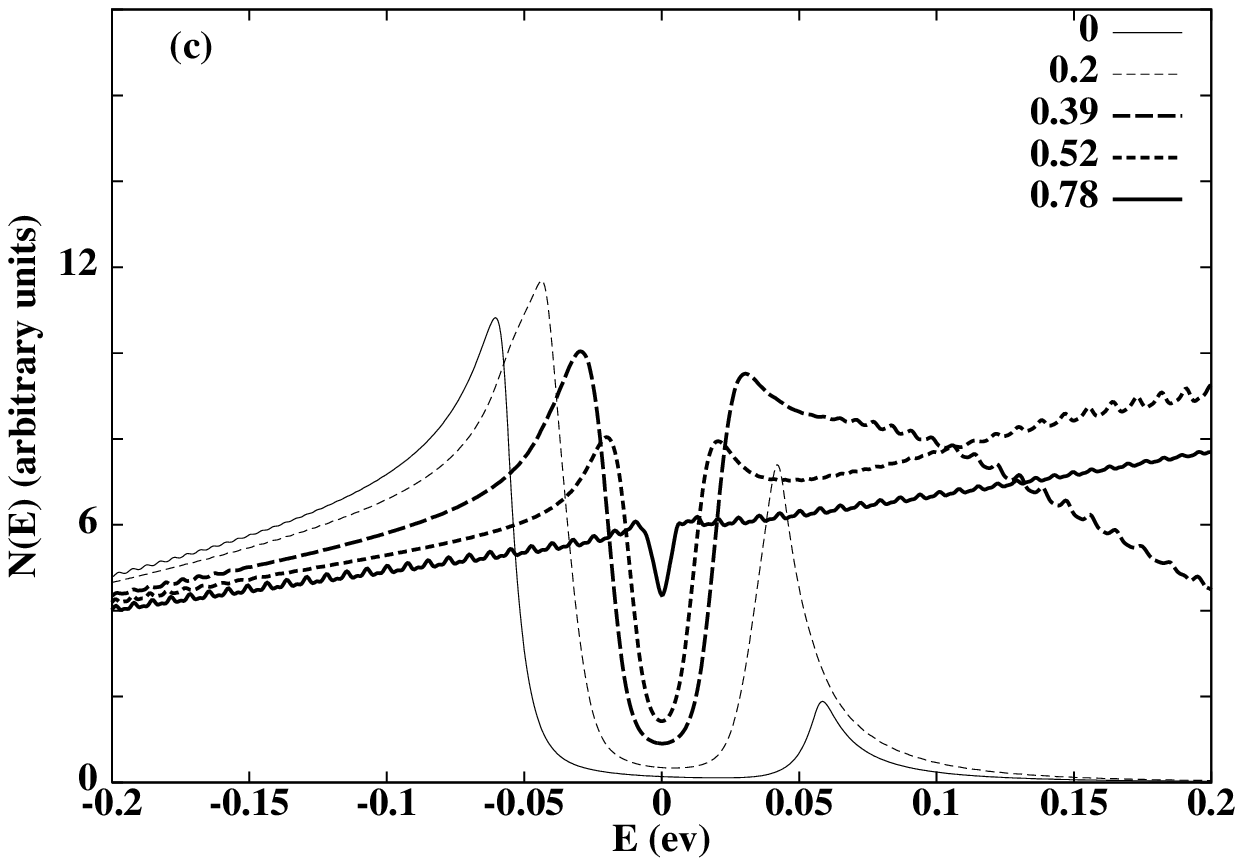}
\hspace{-0.6cm}
\epsfxsize=3.0in
\epsffile{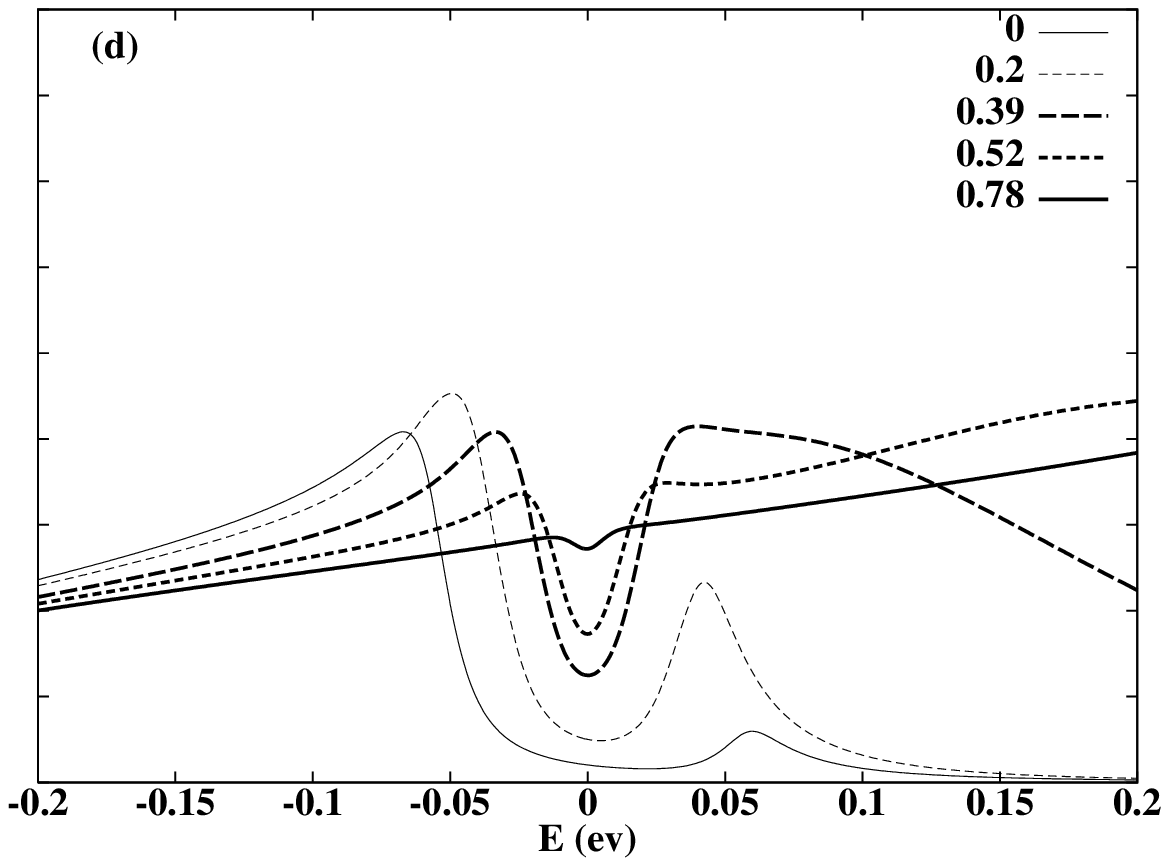}}
\end{center}
\end{figure}
\vspace{-0.6cm}
{\bf Fig.12} Effects of directionality on the DOS as in s-wave (a,b) and d-wave gap symmetry at $\Gamma$=3 meV (a,c) and $\Gamma$=9 meV (b,d).

\begin{multicols}{2}
Fig. 12 has demonstrated this effect. We have presented the DOS at different
$\Theta$ at $\Gamma=3$ meV (a,c) and $\Gamma=9$ meV (b,d) as for s-wave
(a,b) and d-wave (c,d) gap symmetry.
In the case of s-symmetry the position of the quasiparticle peaks is constant excluding the direction
along $k_x$( $\Theta=0$). We pay attention to the strong peaks' asymmetry
in this case.\par
In the case of d-wave symmetry we have practically the same behavior around
$k_x$ direction as for s-wave, but energy gap is changed due to the
$\Theta$-dependence of $\Delta_0$ and correspondingly, the quasiparticle peaks
are shifted to the zero energy.\par
 Fig.13 shows the $\Theta_0$-changing of DOS at  $\Gamma=3$ meV (a,c)
and $\Gamma=9$ meV (b,d) as for s-wave (a,b) and d-wave (c,d) gap symmetry.
The increase of $\Theta_0$ means the taking into play (inclusion) the states,
close to ($\pi,0 $). It is reflected as an appearance of the van Hove singularity
as in case of s-wave and d-wave gap symmetry at small $\Gamma$. The VHS
is more pronounced in case of d-wave in comparing with s-wave symmetry. The increase of $\Gamma$ leads to confluence of the quasiparticle and VHS peaks.\par
We consider that the absence VHS peak on the experimental
$dI/dV$-characteristics means the enough large lifetime broadening factor
$\Gamma$ in that HTSC material.

The origin of the peaks asymmetry in the tunneling DOS was studied in [1]
by considering the role of the tunneling matrix element $|T_k|^2$ in the clean
limit case ($\Gamma=0$), where for the calculation N(E) was used the formula (5).\par
We repeat the explanation of the paper [1] because we believe that the following
conclusion on the origin of the peaks asymmetry must be different.
At $E> 0$ (positive bias voltages) the first term of (5) contributes to the N(E) because of
$\delta(E_k-E)$.
In this case, as can see from Fig. 7 and Fig. 9c and Fig. 14a $|T_k|^2$ selects only a
relatively short region of states in k-space in which $u_k^2>0$.
These are the states with  $\xi_k>0$ (above the FS).
For the majority of states integrated over ( see again Fig.7
and Fig.9c).
At $E< 0$ (negative bias voltages) the second term of (5) contributes to the
DOS because of $\delta(E_k+E)$. In this case, as can see from Fig.7 and
Fig. 9c and Fig. 14b, $|T_k|^2$ selects out a large region of k states where $v_k^2 >0$,
in fact, equal to one. These states are below the Fermi surface, where  $\xi_k<0$.
The overall effect then is to have a large negative bias conductance compared
to the positive one. This is true as for s- and d-wave symmetry. Hence, the
underlying asymmetry of the conductance peaks is primarily due to the band
structure $\xi_k$ and d-wave symmetry simply enhances the degree of asymmetry
of the peaks. So, the peaks' asymmetry existing as for s-wave and d-wave symmetry
is sensitive to
band structure $\xi_k$.
\end{multicols}
\begin{figure}
\begin{center}
\leavevmode
\hbox{%
\epsfxsize=3.0in
\epsffile{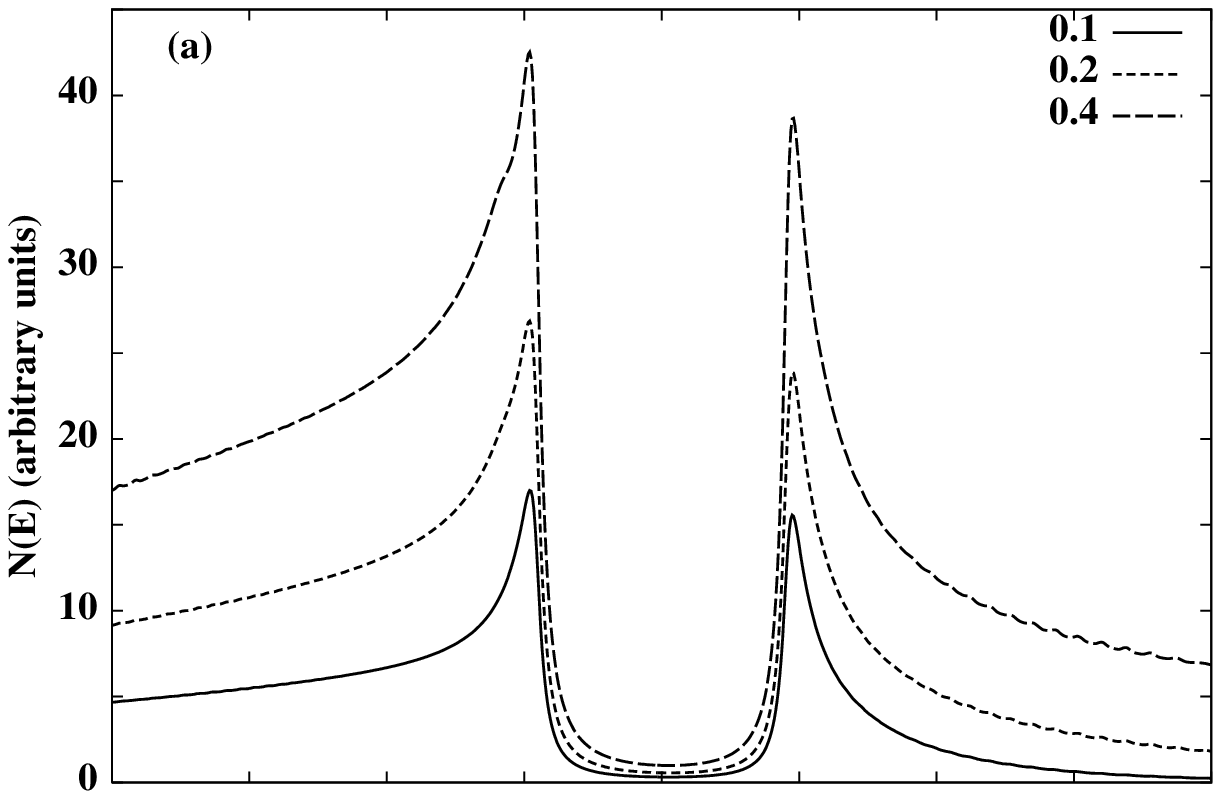}
\hspace{-0.6cm}
\epsfxsize=3.0in
\epsffile{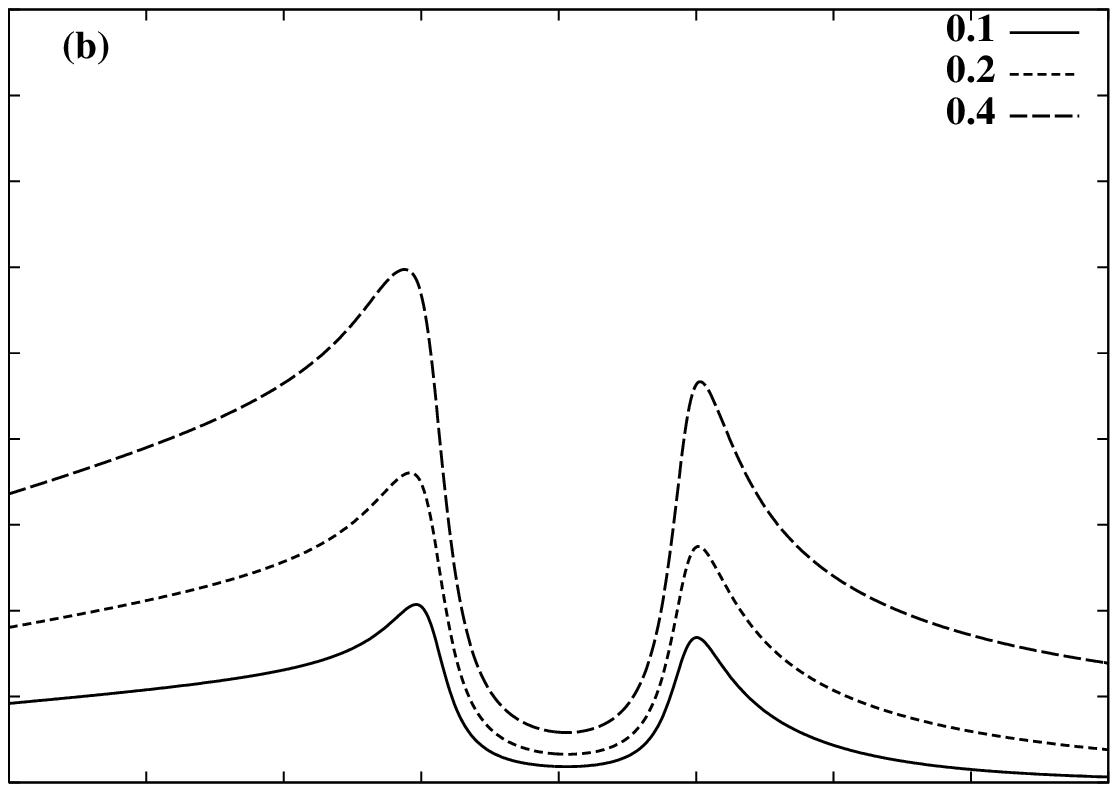}}
\end{center}
\end{figure}
\vspace{-1.78cm}
\begin{figure}
\begin{center}
\leavevmode
\hbox{%
\epsfxsize=3.0in
\epsffile{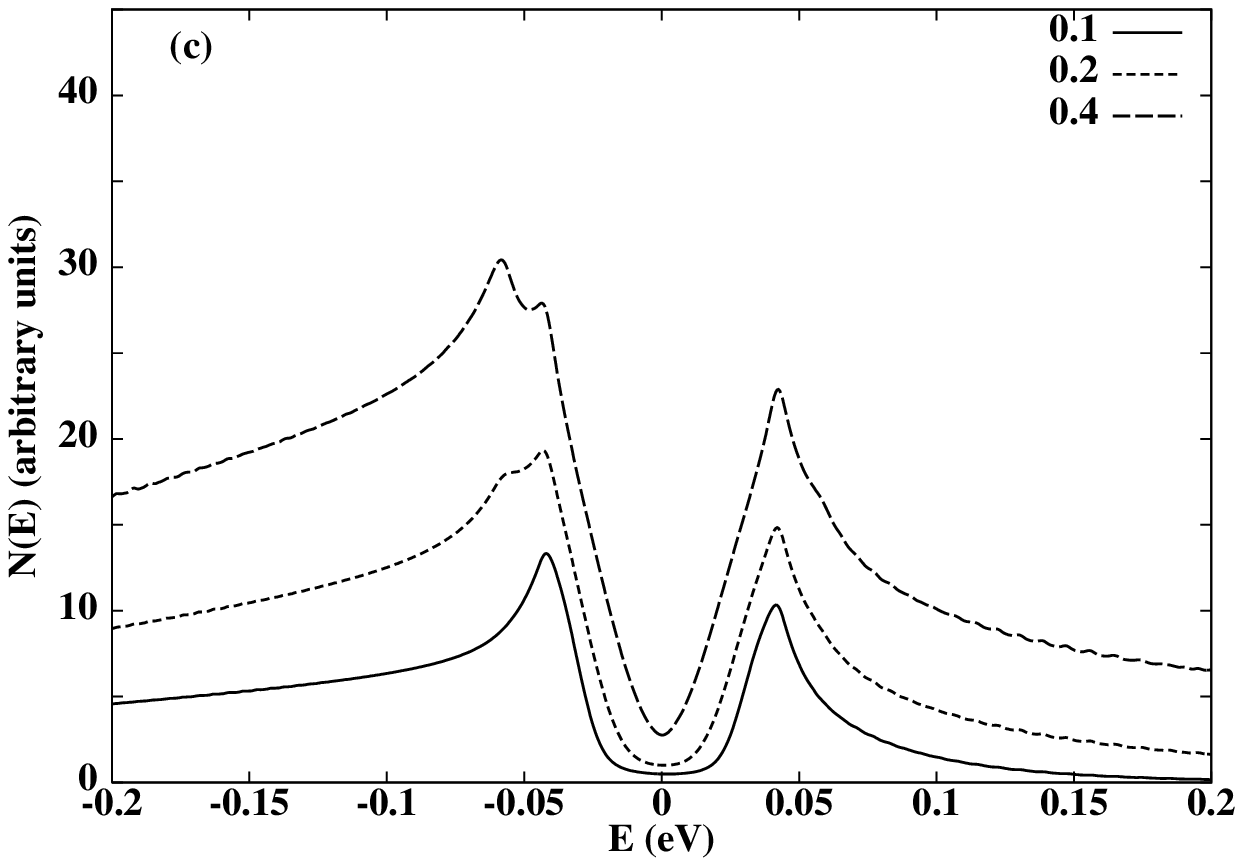}
\hspace{-0.6cm}
\epsfxsize=3.0in
\epsffile{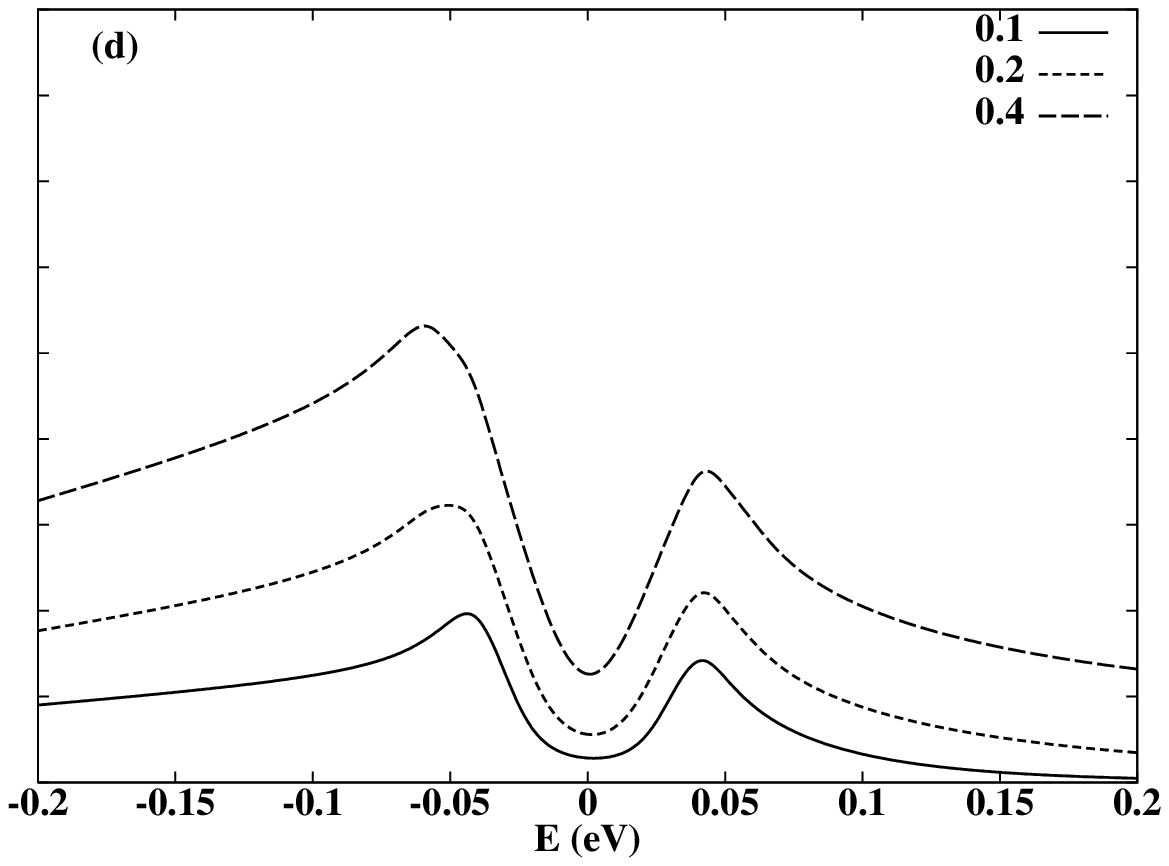}}
\end{center}
\end{figure}
\vspace{-0.6cm}
{\bf Fig.13} Numerical calculation of the quasiparticle DOS with a s-wave (a,b)
and d-wave (c,d)
gap symmetry at $\Gamma$=3 meV (a,c)
 and $\Gamma$=9 meV (b,d)
for different spread $\Theta_0$.
\begin{figure}
\begin{center}
\leavevmode
\hbox{%
\epsfxsize=2.0in
\epsffile{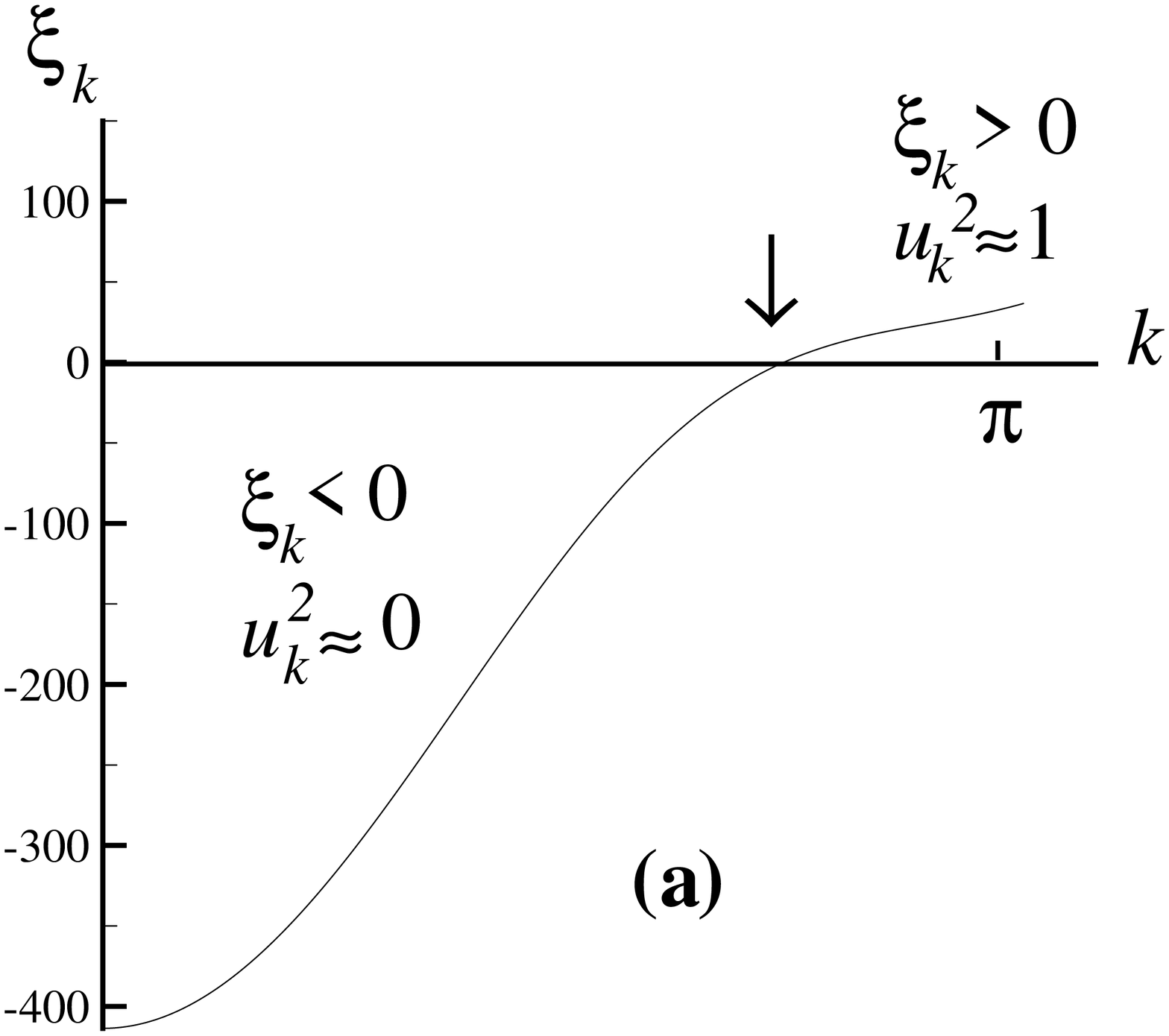}
\hspace{1cm}
\epsfxsize=2.0in
\epsffile{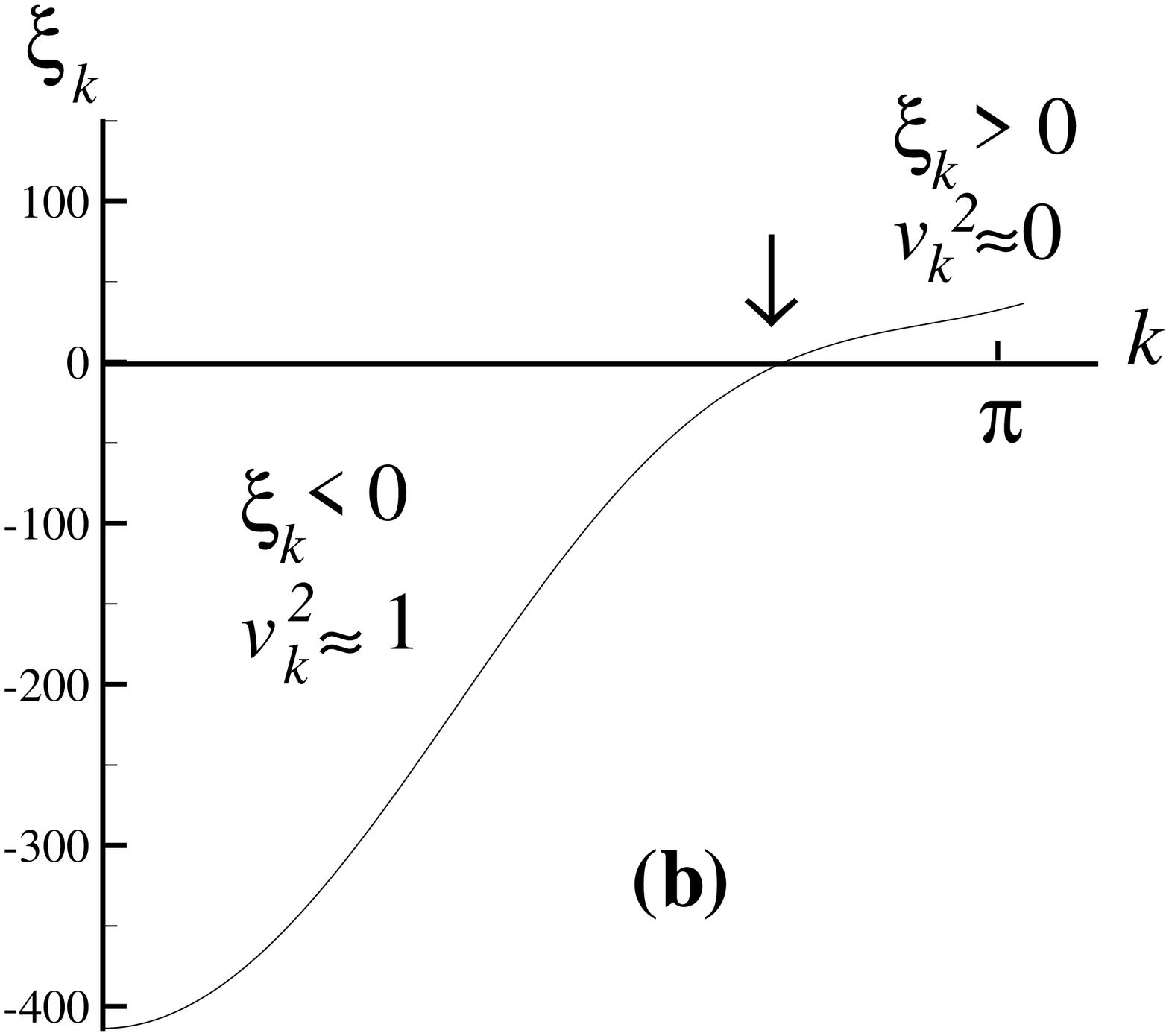}}
\end{center}
\end{figure}
\vspace{-0.6cm}
{\bf Fig.14} ARPES energy spectrum along $\Theta=0.25$. The values of
coherent factors correspond to $E>0$ (a) and $E<0$ (b).
\begin{multicols}{2}
In summary, by changing of the energy gap $\Delta_0$ in HTSC  one may
model the influence of
nonmagnetic impurities on the DOS. We consider that the asymmetry of the
quasiparticle peaks is due to the
specific features of the energy spectrum of HTSC and that the d-wave gap symmetry
only enhances
 the peaks' asymmetry.\par
The absence of the VHS peak on the experimental $dI/dV$-charactristics
means the large enough lifetime
broadening factor $\Gamma$ in HTSC.
\section{Acknowledgement}
We thank Professor Y. Sobouti and Professor M.R. H. Khajehpour for useful discussions.
\end{multicols}
\begin{multicols}{2}

\end{multicols}
\end{document}